\documentclass[twocolumn,preprintnumbers,amsmath,amssymb,showpacs,aps,superscriptaddress]{revtex4}
\usepackage{graphicx}
\usepackage{dcolumn}
\usepackage{bm}
\usepackage{amsmath}
\usepackage{amssymb}
\usepackage{epsf}
\usepackage{epstopdf}
\usepackage{multirow}
\usepackage{color}
\definecolor{Red}{rgb}{1.0,0,0}
\definecolor{Blue}{rgb}{0,0,1.0}
\newcommand{\be}{\begin{equation}}
\newcommand{\ee}{\end{equation}}
\newcommand{\bea}{\begin{eqnarray}}
\newcommand{\eea}{\end{eqnarray}}

\newcommand{\upstate}{|\!\!\uparrow\rangle}
\newcommand{\downstate}{|\!\!\downarrow\rangle}

\begin{document}
\title{Spectroscopy and Thermometry of Drumhead Modes in a Mesoscopic Trapped-Ion Crystal using Entanglement}
\author{Brian C. Sawyer}
\email{brian.sawyer@boulder.nist.gov}
\author{Joseph W. Britton}
\affiliation{Time and Frequency Division, National Institute of Standards and Technology, Boulder, CO  80305}
\author{Adam C. Keith}
\altaffiliation{Department of Physics, North Carolina State University, Raleigh, NC 27695}
\author{C.-C. Joseph Wang}
\author{James K. Freericks}
\affiliation{Department of Physics, Georgetown University, Washington, DC 20057}
\author{Hermann Uys}
\affiliation{Council for Scientific and Industrial Research, Pretoria, South Africa}
\author{Michael J. Biercuk}
\affiliation{Centre for Engineering Quantum Systems, School of Physics, The University of Sydney, NSW Australia}
\author{John J. Bollinger}
\affiliation{Time and Frequency Division, National Institute of Standards and Technology, Boulder, CO  80305}

\begin{abstract}
We demonstrate spectroscopy and thermometry of individual motional modes in a mesoscopic 2D ion array using entanglement-induced decoherence as a method of transduction. Our system is a $\sim$400 $\mu$m-diameter planar crystal of several hundred $^9$Be$^+$ ions exhibiting complex drumhead modes in the confining potential of a Penning trap. Exploiting precise control over the $^9$Be$^+$ valence electron spins, we apply a homogeneous spin-dependent optical dipole force to excite arbitrary transverse modes with an effective wavelength approaching the interparticle spacing ($\sim$20 \nolinebreak$\mu$m). Center-of-mass displacements below 1 nm are detected via entanglement of spin and motional degrees of freedom.
\end{abstract}


\maketitle

Studies of quantum physics at the interface of microscopic and mesoscopic regimes have recently focused on the observation of quantum coherent phenomena in optomechanical systems~\cite{OConnell10,Kippenberg08,Teufel09}. The realization of quantum coherence in mechanical oscillations involving many particles behaving approximately as a continuum provides exciting insights into the quantum-classical transition. Previous work has shown that crystals of cold, trapped ions behave as atomic-scale nanomechanical oscillators~\cite{Biercuk10,Jost09,Brown11}, with the benefits of in-situ tunable motional modes and exploitable single-particle quantum degrees of freedom (e.g. valence electron spin). Our system of hundreds of crystallized ions in a Penning trap provides a bottom-up approach to studying mesoscopic quantum coherence. In this context, the relevant particle numbers are sufficiently small to permit excellent quantum control without sacrificing continuum mechanical features.  Beyond these capabilites, trapped ions have long provided a laboratory platform for studying diverse physical phenomena including: strongly-coupled one-component plasmas (OCPs)~\cite{Ichimaru82,Dubin99}; quantum computation~\cite{Hanneke09,Monz09} and simulation~\cite{Friedenauer08,Kim10,Islam11,Britton11,Lanyon11}; dynamical decoupling~\cite{Biercuk09nature}; and atomic clocks and precision measurement~\cite{Rosenband08}.

\begin{figure}[b!]
\resizebox{8.7cm}{!}{
\includegraphics{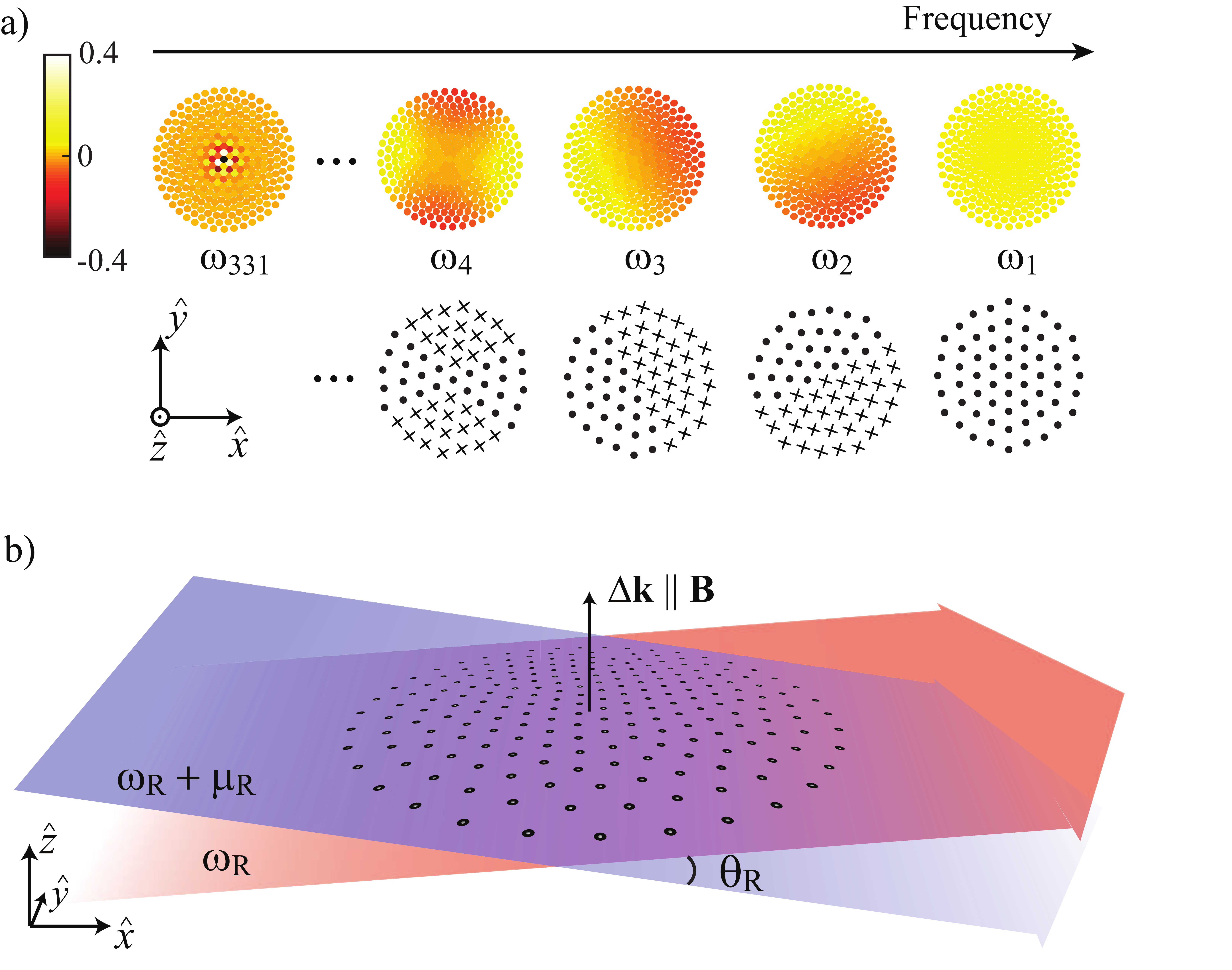}}
\caption{\label{fig1}(color online) (a) Calculated structure of selected transverse eigenmodes ($\vec{b}_m$) for a 2D crystal of 331 $^9$Be$^+$ ions. Mode frequencies, $\omega_m$, decrease as effective wavelength gets shorter. The arbitrary color scale indicates relative ion displacement amplitude. One example of an ion spin state with similar symmetry is given below each of the four highest-frequency eigenmodes. The symbol $\times$($\bullet$) denotes spin-projection into (out of) the plane. Interaction between these spin and mode configurations mediated by the spin-dependent optical dipole force (ODF) leads to excitation of the corresponding eigenmode. (b) Illustration of a single plane of $^9$Be$^+$ within the Penning trap. Two 313-nm beams intersect at the ion cloud to form a traveling wave of beat frequency $\mu_R$ and effective wavevector $\protect\overrightarrow{\Delta k}$ along the direction of the trap magnetic field. The electric field intensity is uniform in the plane, but the spin-dependent induced AC Stark shift permits excitation of transverse modes of arbitrary wavelength.}
\end{figure}

In this Letter, we present an experimental and theoretical study of motional drumhead modes in a 2D crystal of $^9$Be$^+$ ions confined within a Penning trap. We excite \emph{inhomogeneous} modes of arbitrary wavelength (see Fig.~\ref{fig1}(a)) through application of a \emph{homogeneous}, spin-state-dependent optical dipole force (ODF) to a large-scale spin superposition. Distinct drumhead modes are entangled with the $^9$Be$^+$ valence electron spins by tuning a beat frequency ($\mu_R$) between two ODF lasers near a mode resonance. This spin-motion entanglement is detected as a $\mu_R$-dependent decoherence of ion spins whose magnitude conveys the specific mode temperature.

\newpage
Previous global mode studies on 2D planar ion arrays were restricted to modes with wavelengths on the order of the cloud size~\cite{Heinzen91,Bollinger93,Weimer94,Tinkle94,Dantan10}. By contrast, the short-wavelength modes studied here are of particular interest due to their increased sensitivity to strong-correlation corrections~\cite{Kriesel02,Castro10} compared to those with long wavelength, which are well-described by fluid theory. Thermometry of large Coulomb crystals has thus far been limited to determination of global temperature through Doppler profile measurements~\cite{Jensen04}, which give a minimum sensitivity of $\sim$0.5 mK in $^9$Be$^+$. Our temperature measurement is mode-specific and may be employed below the Doppler cooling limit, providing an alternative to Raman sideband thermometry~\cite{Monroe95}.

The Penning trap used for this work is detailed in a previous publication~\cite{Biercuk09}. Application of static voltages to a stack of cylindrical electrodes provides harmonic confinement along $\hat{z}$ (the trap symmetry axis) with a $^9$Be$^+$ center-of-mass (COM) oscillation frequency of $\omega_1/2\pi = 795$ kHz that is independent of the number of trapped ions. The trap resides within the room-temperature bore of a superconducting magnet, and radial confinement is achieved via the Lorentz force generated by rotation of the ion cloud through the static, homogeneous magnetic ($B$) field of $\sim$4.46 T oriented along $\hat{z}$. Application of a time-dependent quadrupole `rotating wall' potential permits phase-stable control of the rotation frequency ($\omega_r$), and thus the confining radial force of the trap~\cite{Hasegawa05,Huang98}. In the limit of a weak rotating wall potential, the harmonic trap potential in a frame rotating at $\omega_r$ is~\cite{Dubin99}
\begin{eqnarray}
q\Phi_{\text{trap}}(r,z) &=& \frac{1}{2}M\omega_1^2 \left(z^2 + \beta r^2 \right), \\
\beta &=& \frac{\omega_r(\Omega_c - \omega_r)}{\omega_1^2} - \frac{1}{2} \label{beta}
\end{eqnarray}
where $M$ ($q$) is the mass (charge) of a single $^9$Be$^+$, $\Omega_c=2\pi \times 7.6$ MHz is the cyclotron frequency, and $z$ ($r$) is axial (radial) distance from the trap center. We set the rotation frequency, $\omega_r$, such that the radial confinement is weak relative to transverse confinement ($\beta\ll1$), resulting in a single ion plane.

\begin{figure}[t]
\resizebox{8.3cm}{!}{
\includegraphics{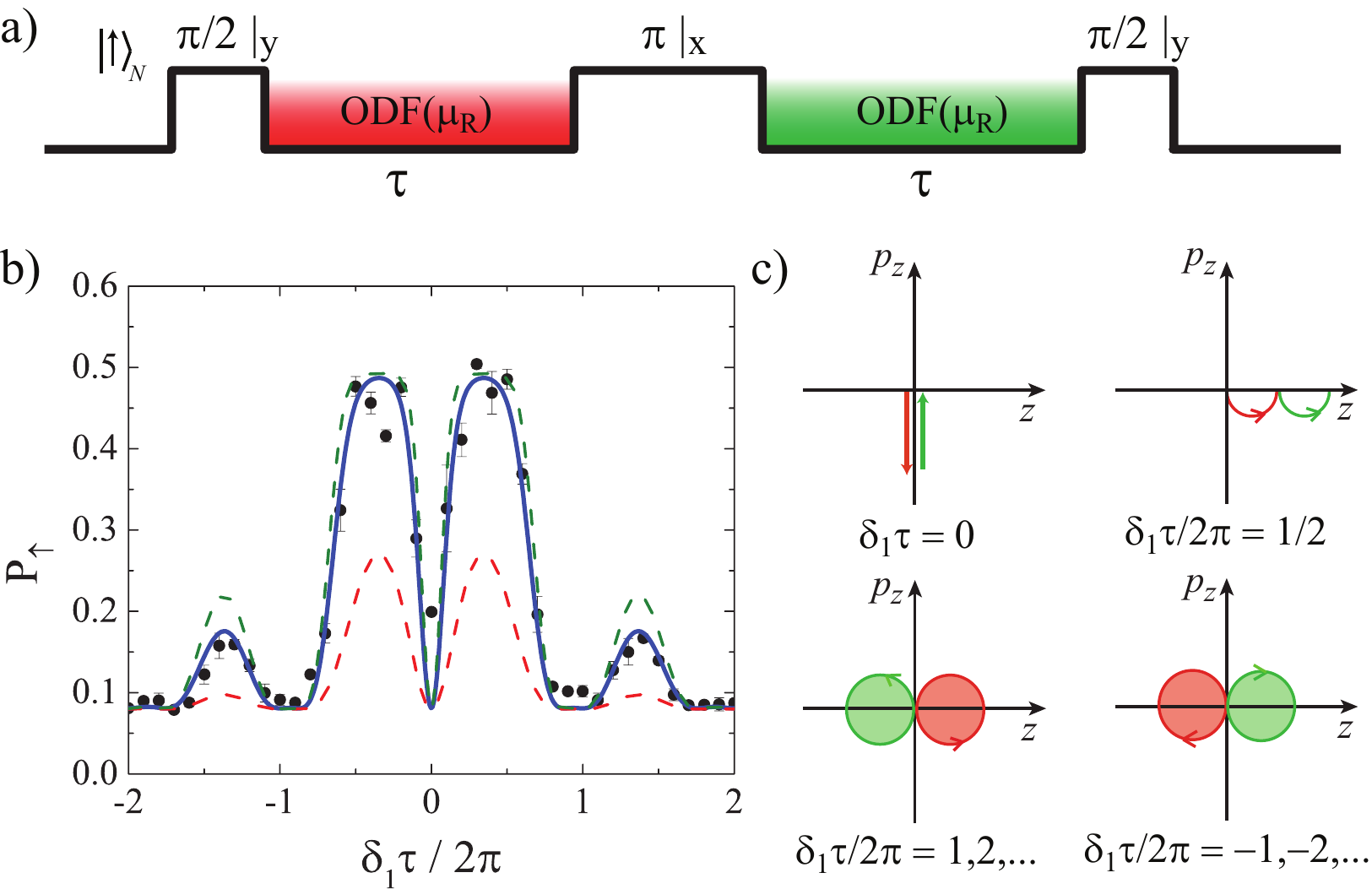}}
\caption{\label{fig2}(color online) (a) Pulse sequence used for excitation and detection of transverse motional modes. Global spin rotations are performed with microwaves at $\sim$124 GHz, while the state-dependent optical dipole force is applied in each arm of the spin echo for a duration $\tau$. We implement $\pi$-pulse times ($t_{\pi}$) as short as 65 $\mu$s. (b) Measured (points with statistical error bars) and fit (solid blue line) probability of detecting $\upstate$ ($P_{\uparrow}$) at the end of the spin echo sequence. Frequency-dependent decoherence is due to entanglement of spins with the axial COM mode ($\omega_1/2\pi=795$ kHz) as a function of ODF detuning $\delta_1\equiv\mu_R - \omega_1$ in a cloud of $190\pm8$ ions. Each point is an average of 90 experimental runs. The fit provides a mode temperature of $2.3\pm 0.5$ mK, whose error includes a 5\% uncertainty in ODF beam angle, $\theta_{\text{R}}$. For comparison, the lower (upper) dashed line is calculated assuming 0.4 mK (4.0 mK). (c) Illustrated phase-space trajectories of state $\upstate_N$ at different detunings, $\delta_1$, in a frame rotating at $\omega_1$. Axis labels represent COM momentum ($p_z \propto \text{Im}[\alpha_{j1}]$) and position ($z \propto \text{Re}[\alpha_{j1}]$).}
\end{figure}

The $m_J=\pm1/2$ projections of the Be$^+$ $^2S_{1/2}$ ground state are split by $\sim$124 \nolinebreak GHz and serve as $\upstate$ and $\downstate$ `qubit' states, respectively. Global spin rotations are performed by injecting 124-GHz radiation through a waveguide attached to the side of the trap. The $^9$Be$^+$ ions are Doppler laser cooled with laser beams directed both parallel and perpendicular to $\hat{z}$. Both beams are tuned to the \nolinebreak$^2S_{1/2}(m_J=+1/2)$--\nolinebreak$^2P_{3/2}(m_J=+3/2)$ transition at $\sim$313 nm to cool ion motion below 1 mK. This same transition is used for ion detection and projective spin-state measurement. Discrimination of $\upstate$ (bright) from $\downstate$ (dark) is performed with a fidelity $>99$\%~\cite{Biercuk09}.

The axial and radial confining potentials are tuned to yield a planar ion configuration. Due to mutual Coulomb repulsion and the low ion temperature, the ions' minimum-energy configuration is a 2D crystal with triangular order~\cite{Mitchell98}. Ion spacing is $\sim$20 $\mu$m, and individual ions can be resolved using stroboscopic imaging at $\omega_r$. The planar array of $N$ ions exhibits $3N$ motional modes, $N$ of which are drumhead oscillations transverse to the crystal plane (see Fig.~\ref{fig1}(a)). As with 1D ion strings, the frequencies of these transverse modes decrease with decreasing effective wavelength due to screening of confining electric fields by nearby ions. The transverse eigenvectors ($\vec{b}_m$, $m\in [1,N]$) and corresponding eigenfrequencies ($\omega_m$) are obtained by first numerically calculating the zero-temperature 2D ion configuration in the presence of the Penning trap potentials. Applying a Taylor expansion of the combined trap and Coulomb potential about each ion equilibrium position, we diagonalize the $N\times N$ stiffness matrix whose eigenvalues and unit eigenvectors are $\omega_m$ and $\vec{b}_m$, respectively~\cite{Zhu06,Kim09}. The relative displacement amplitude of an ion $j$ is given by the $j$th element of $\vec{b}_m$, denoted as $b_{jm}$, where $\sum_m \left|b_{jm}\right|^2 = \sum_j \left|b_{jm}\right|^2 = 1$.

To excite transverse modes in our 2D Coulomb crystal, we employ a spin-dependent ODF generated by interfering two off-resonant laser beams at the ion cloud position. This is depicted schematically in Fig.~\ref{fig1}(b). The two ODF beams are produced from a single beam using a 50/50 beamsplitter and subsequently pass through separate acousto-optic modulators that allow fast ($\sim$1 $\mu$s) switching and impart a relative detuning $\mu_R$. The beams intersect at an angle of $\theta_{\text{R}}=4.8^{\circ}\pm 0.2^{\circ}$ at the ion cloud position, and their relative alignment is adjusted to orient the effective wavevector ($\overrightarrow{\Delta k}$) of the resulting standing ($\mu_R=0$) or traveling ($\mu_R\neq0$) wave to within $\sim0.05^{\circ}$ of $\hat{z}$. The common wavelength (313.133 nm) and unique linear polarizations of the beams are chosen such that the AC Stark shift from the interfering beams on state $\upstate$ is equal in magnitude and opposite in sign to that on $\downstate$~\cite{EPAPS}. The result of the interference between these two beams is a spin-dependent force on each ion, $j$ ($F_{\uparrow,j}=-F_{\downarrow,j}\equiv F_j$). The Hamiltonian for this interaction is $\hat{H}_{\text{ODF}}=-\sum_{j=1}^{N}F_j \hat{z}_j(t)\cos{(\mu_R t)}\hat{\sigma}^z_j$, where $\hat{z}_j(t)$ is the time-dependent position operator and $\hat{\sigma}^z_j$ is the $z$-component Pauli operator for ion $j$~\cite{Britton11}. The elliptical beam waists (100 $\mu\text{m}\times1000$ $\mu$m, with the major axis oriented parallel to the ion plane~\cite{EPAPS}) are sufficiently large to generate an approximately uniform ODF with variation below 10\% across the $\sim$400 $\mu$m-diameter planar ion crystal. Typical ODFs for this work are $F_j\sim10^{-23}$ N along $\hat{z}$.

Figure~\ref{fig2}(a) illustrates the experimental control sequence for microwaves (black line) and ODF lasers (shaded regions) used to coherently excite transverse modes of motion. Ions are first prepared in the `bright' state $\upstate_N\equiv\prod_{j=1}^N | \!\!\uparrow_j\rangle$ via optical pumping~\cite{Biercuk09}. The sequence of microwave pulses in Fig.~\ref{fig2}(a) comprises a spin echo (SE)~\cite{Hahn50} that, in the absence of the ODF beams, rotates the ions to the `dark' state $\downstate_N$ with $>$99\% fidelity. The SE cancels low-frequency precession about $\hat{z}$ due to ODF laser intensity and magnetic field fluctuations as well as microwave phase noise~\cite{Uys09,Biercuk09nature}. The spin-dependent ODF is applied in each arm of the SE for a duration $\tau$.

The initial microwave pulse rotates each spin by $\pi/2$ to produce the state $\prod_{j=1}^N \frac{1}{\sqrt{2}}\left( | \!\! \uparrow_j\rangle - | \!\!\downarrow_j\rangle\right)$,
which is a superposition of all possible ($2^N$) spin permutations. Importantly, it is the creation of this state that permits subsequent excitation of arbitrary transverse modes with our homogeneous, spin-dependent ODF. By tuning $\mu_R$ near a mode of frequency $\omega_m$, the spin-dependent ODF excites those components of the spin superposition with approximately the same symmetry as the eigenvector $\vec{b}_m$. A subset of these eigenvectors and associated spin states are illustrated in Fig.~\ref{fig1}(a). Depending on experimental parameters, the spin states may be entangled with different motional states at the end of the control sequence of Fig.~\ref{fig2}(a). Upon measurement of the spin state (performing a trace over the motion), entanglement is manifested as spin decoherence that varies with $\mu_R$. We observe this as a decrease in the length of the spins' Bloch vector and a concomitant increase in the probability ($P_{\uparrow}$) of measuring state $\upstate$ averaged over all ions.

Figure~\ref{fig2}(b) gives experimental and theoretical results for a sweep of $\mu_R$ near the COM frequency, $\omega_1$, with $\tau = 500$ $\mu$s and $\delta_1 = (\mu_R - \omega_1)$. On resonance ($\delta_1 = 0$), the pulse sequence leads to excitation (de-excitation) of the COM mode in the first (second) arm. When the product $|\delta_1 \tau/2\pi|=l$ is a non-zero integer, each spin state traverses $l$ full loops in phase space over $\tau$ (see Fig.~\ref{fig2}(c)). At intermediate detunings, the spin and motion remain entangled at the end of the pulse sequence, producing the lineshape of Fig.~\ref{fig2}(b). These motional excitations are described by the spin-dependent displacement operator $\hat{U}(\tau) = \prod_{j,m} \exp{\left[(\alpha_{jm}\hat{a}^{\dag}_{m}-\alpha^{\ast}_{jm}\hat{a}_{m})\hat{\sigma}^z_{j}\right]}$~\cite{Kim09,Monroe96,EPAPS}, where $\alpha_{jm}(\tau)$ is the coherently-driven complex displacement amplitude for ion $j$ of mode $m$, and $\hat{a}^{\dag}_m (\hat{a}_m)$ is the creation (annihilation) operator for mode $m$. Accounting for both arms of the pulse sequence, we obtain~\cite{EPAPS}
\begin{widetext}
{\footnotesize
\begin{equation}\label{alpha}
\alpha_{jm} = \frac{F_{j} b_{jm}}{\hslash(\mu_R ^2-\omega_m ^2)}\sqrt{\frac{\hslash}{2M\omega_m}} \left[\omega_m(1-\cos{\phi})+i\mu_R\sin{\phi}-e^{i\omega_m \tau} \left\{ \omega_m \left[\cos{(\mu_R\tau)}-\cos{(\mu_R\tau+\phi)}\right] - i\mu_R\left[\sin{(\mu_R\tau)-\sin{(\mu_R\tau+\phi)}}\right] \right\} \right],
\end{equation}
}
\end{widetext}
where $\hslash$ is Planck's constant, $F_j$ is the ODF magnitude on ion $j$, and $\phi=(\tau+t_{\pi})(\mu_R - \omega_m)$ accounts for phase evolution of the ODF drive relative to that of the mode.

\begin{figure*}[t]
\resizebox{16cm}{!}{
\includegraphics{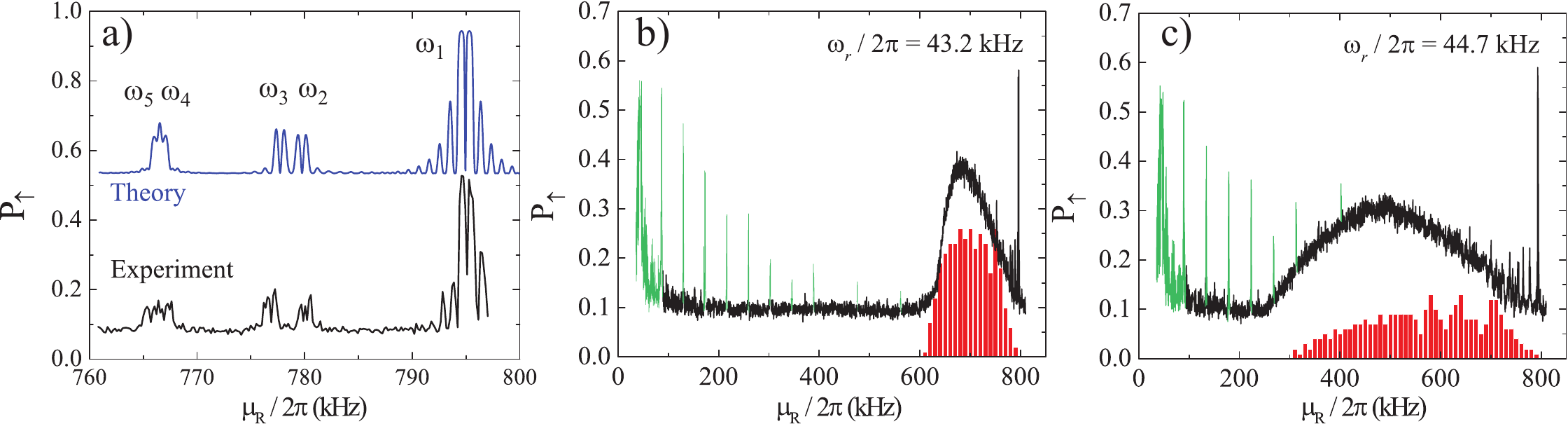}}
\caption{\label{fig3}(color online) (a) Measured (lower) and calculated (offset) probabilities for measuring $\upstate$ after the spin echo sequence as a function of ODF beat frequency for a sweep of $\mu_R$ over the first five transverse modes with $250\pm15$ ions. The modes at $\omega_2$ and $\omega_3$ are split due to distortion of the ion cloud boundary by the rotating wall potential. Panels (b) and (c) give results of wider sweeps with $\omega_r/2\pi=$ 43.2 kHz and 44.7 kHz, respectively, in a crystal of $345\pm25$ ions. Frequency-dependent deviation from P$_{\uparrow}\sim0.1$ is due to spin-motional entanglement, while the background is due to spontaneous emission from the ODF beams. The histogram (red bars) shown below each experimental curve depicts the density of calculated eigenmodes at the given $\omega_r$. Histogram bins are 10 kHz wide and plotted with an arbitrary vertical scale. As described in Fig.~\ref{fig1}(a), the highest-frequency feature is that of the COM mode and the $\sim$50 lowest-frequency eigenmodes include nearest-neighbor ions oscillating out of phase. Features at $\omega_r$ and precise harmonics thereof (shaded in light green) are due to spin-motion entanglement with in-plane degrees of freedom excited by the small ($\sim10^{-3}F_j$) component of ODF perpendicular to $\hat{z}$.}
\end{figure*}

Although the coherently driven, spin-dependent displacements ($\alpha_{jm}$) are independent of the initial motional state (assuming Lamb-Dicke confinement~\cite{Britton11}), the spin-motion entanglement signal in Fig.~\ref{fig2}(b) sensitively depends on this initial state. This can be qualitatively understood in terms of the spatial structure of a harmonic oscillator Fock state, $|n_m\rangle$, of mode $m$. A state $|n_m\rangle$ exhibits $n$ wavefunction nodes and therefore, as $n$ increases, a fixed spin-dependent displacement results in less wavefunction overlap between different spin components due to the increasing spatial frequency of $|n_m\rangle$ wavefunctions. This leads to larger decoherence and greater displacement sensitivity as the average mode occupation, $\bar{n}_m$, is increased for a given mode. We fit the experimental measurements in Fig.~\ref{fig2}(b) using theory that attributes a thermal state of motion to each mode $m$ characterized by mode occupation $\bar{n}_m \sim k_B T_m(\hslash \omega_m)^{-1}$ and temperature $T_m$. Neglecting spin-spin correlation contributions, we find the probability $P^{(j)}_{\uparrow}$ of detecting ion $j$ in state $\upstate$ at the end of the pulse sequence to be~\cite{EPAPS}
{\footnotesize
\begin{equation}\label{Pup}
P_{\uparrow}^{(j)} = \!\! \frac{1}{2}\left[1-e^{-2\Gamma\tau}\exp{\left(-2\sum_m |\alpha_{jm}|^2 (2\bar{n}_m+1)\right)} \right].
\end{equation}}
Here $\Gamma$ accounts for decoherence due to spontaneous emission induced by the ODF lasers over the duration $2\tau$, and is responsible for the background level of $P_{\uparrow}\sim0.1$ observed in all experimental data presented here~\cite{Uys10}. The total detection probability $P_{\uparrow}$ is obtained by averaging all $P_{\uparrow}^{(j)}$.

For interaction with the COM mode ($b_{j1}=\frac{1}{\sqrt{N}}, \forall j \in [1,N]$), $\alpha_{j1}$ is obtained from Eq.~(\ref{alpha}) through measurement of the ODF laser intensities~\cite{Britton11} and trapped-ion number, while $\Gamma$ is determined from decoherence observed with $\mu_R$ detuned far from any modes. As such, the only parameter of Eq.~(\ref{Pup}) not measured directly is $\bar{n}_1$, which is varied to fit experimental data as in Fig.~\ref{fig2}(b), where we obtain $\bar{n}_1=60\pm13$ ($T_1 = 2.3\pm0.5$ mK).

We note that a detectable phase-space displacement is obtained with a very small amplitude of $|\alpha_{jm}|$. For example, in Fig.~\ref{fig2}(b), the 20\% decrease in the Bloch vector at $\delta_1\tau/2\pi \backsimeq \pm 1.4$ corresponds to a spin-state-dependent excitation of the COM mode with a mean excursion of $\sim$0.6 nm in each arm of the pulse sequence. This shift is less than 0.2\% of the wavefunction spread of a single ion in the planar array. Our sensitivity to displacements improves with increasing mode temperature provided that the ODF is adjusted to avoid full decoherence ($P_{\uparrow}=0.5$) at the detuning of interest.

Figure~\ref{fig3}(a) shows the result of a sweep of $\mu_R$ over five transverse modes and corresponding theory. The theoretical spectrum (offset for clarity) is generated assuming $T_1=10$ mK and $T_{m>1}=0.4$ mK, with $T_1$ obtained from a fit. The large COM temperature of Fig.~\ref{fig3}(a) is produced by quickly switching off the $\hat{z}$-oriented Doppler cooling beam on a time scale of $\sim$$2\pi\omega_1^{-1}$. In this case, sudden loss of radiation pressure from the cooling light induces a COM oscillation amplitude of $\sim$50 nm that we detect as an elevated $\bar{n}_1$. A more adiabatic reduction of the cooling beam intensity yields $\bar{n}_1\sim26$ ($T_1\sim1$ mK). For modes other than the COM, we must additionally calculate the $b_{jm}$ values for the trap potentials and ion number in a given experiment. For these modes, we find temperatures consistent with the Doppler cooling limit of 0.43 mK.

To measure the full spectrum of transverse modes, we repeat the sequence of Fig.~\ref{fig2}(a) for $30 \text{ kHz} \leq \mu_R/2\pi \leq 800 \text{ kHz}$ with $\tau=1$ ms. With the exception of the COM mode, the frequencies of the remaining $N-1$ modes depend sensitively on our choice of crystal rotation frequency, $\omega_r$~\cite{Weimer94}. Figures~\ref{fig3}(b)-(c) show the result of these experimental runs for $\omega_r/2\pi=43.2$ kHz and 44.7 kHz, respectively. For this ion number of $345\pm25$, the single-plane configuration is stable over the range $42.2~\text{kHz}\lesssim\omega_r/2\pi \lesssim 45.2~\text{kHz}$. Histograms of calculated mode density versus $\mu_R/2\pi$ are plotted below each experimental curve with an arbitrary vertical scale and bin width of 10 kHz. The distribution of eigenfrequencies narrows as $\omega_r$ is decreased; weaker radial confinement (see Eq.~(\ref{beta})) leads to lower ion densities and reduced screening of trap potentials, thereby moving the frequency of the shortest-wavelength mode toward that of the COM. This behavior is clearly visible in Figs.~\ref{fig3}(b)-(c). Additionally, we find quantitative agreement between the measured spectrum and that generated from numerical calculation of the transverse eigenmodes under the given experimental conditions, documenting coupling to both short- and long-wavelength modes. The sharp features of Figs.~\ref{fig3}(b)-(c) shaded in light green reflect excitation of in-plane resonances at harmonics of $\omega_r$ due to a very small component of the ODF ($\sim$$10^{-3}F_j$) along the ion plane. These spectral features may be reduced through more careful alignment of $\overrightarrow{\Delta k}$ to $\hat{z}$, but their strong response suggests an elevated motional temperature perpendicular to $\hat{z}$.

In summary, we have used entanglement of spin and motional degrees of freedom to map the full transverse mode spectrum of a mesoscopic 2D ion array. This technique provides a tool for sensitively and accurately measuring the temperature and displacement amplitude of individual drumhead modes, facilitating identification of mode-specific heating mechanisms and the resulting non-equilibrium energy distributions. Coherent, spin-dependent excitation of transverse modes is the basis for engineering quantum spin-spin interactions with trapped ions~\cite{Friedenauer08,Kim10,Zhu06,Porras04,Porras06,Islam11,Britton11,Kim09}, making mode characterization a critical element of such experiments. Future work will include investigation of low-frequency in-plane modes at frequencies smaller than $\omega_r$. A predicted subset of these modes includes in-plane shearing motion whose restoring force is due exclusively to strong correlations.

This work was supported by the DARPA-OLE program and NIST. B. C. Sawyer is supported by a NRC fellowship funded
by NIST. M. J. Biercuk and J. J. Bollinger acknowledge partial support from the ARC Centre of Excellence for Engineered Quantum Systems, CE110001013. A. C. Keith was supported by the NSF under grant number DMR-1004268. J. K. Freericks was supported by the McDevitt endowment bequest at Georgetown University. We thank D.H.E. Dubin, D. Porras, K.-K. Ni, D. Slichter, and S. Manmana for comments on the manuscript. This manuscript is a contribution of NIST and not subject to U.S. copyright.

\newpage
\onecolumngrid

\begin{center}
\textbf{\underline{Supplementary Material}:\\ Spectroscopy and Thermometry of Drumhead Modes in a Mesoscopic Trapped-Ion Crystal using Entanglement}
\end{center}

\begin{figure}[b]
\includegraphics[width=12cm]{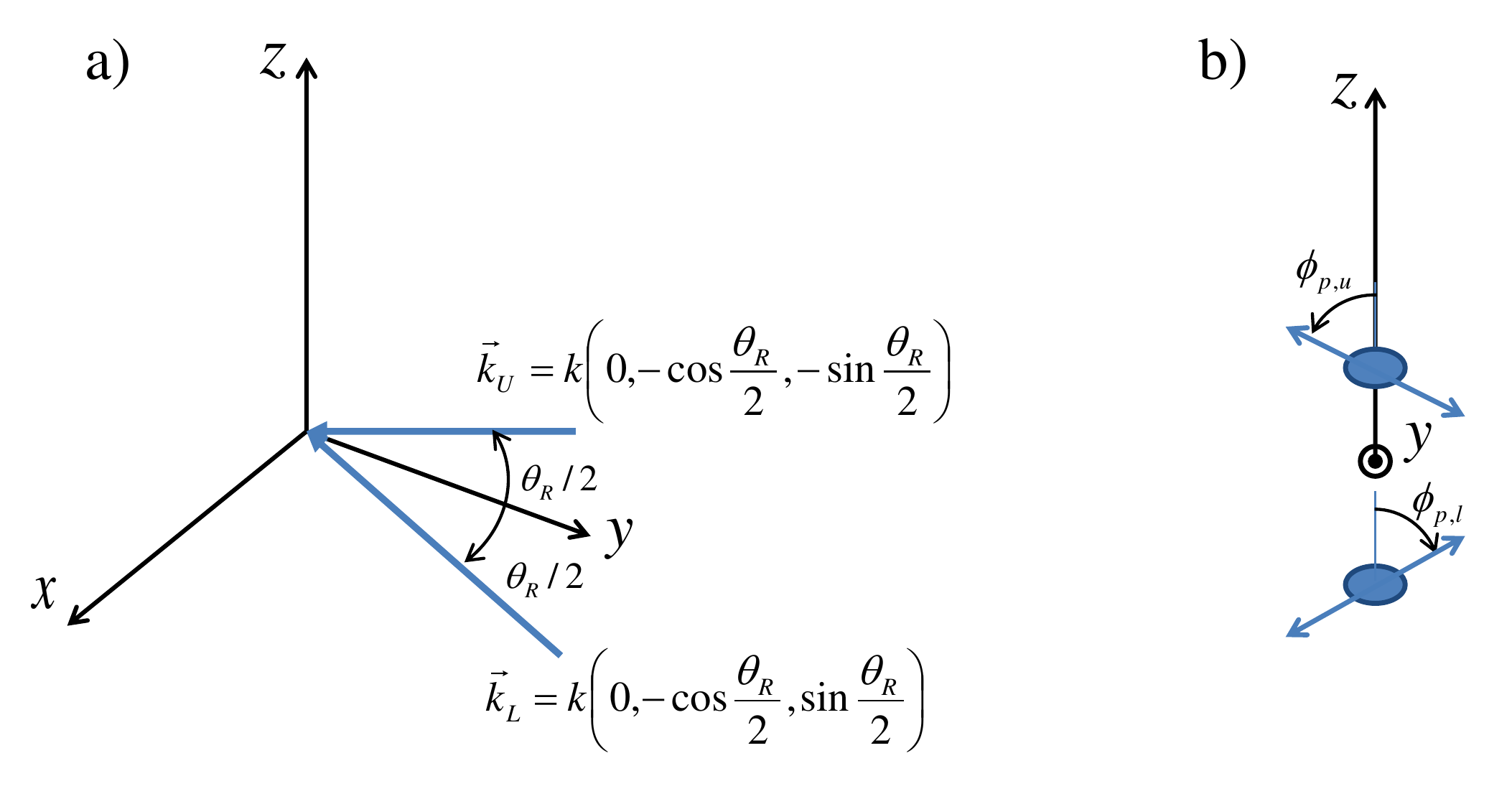}

\caption{Sketch of ODF laser beam setup. a) The ODF laser beams lie in the
$y$-$z$ plane at angles $\pm\theta_{R}/2$ with respect to the $y$-axis.
b) View looking in the $-\hat{y}$ direction. The beams are linearly
polarized but with different polarization angles relative to vertical
polarization.\label{fig:ODF_setup}}
\end{figure}

\section*{Optical Dipole Force Details}

Figure~\ref{fig:ODF_setup} shows a simple sketch of the optical dipole
force (ODF) laser beam set-up. As discussed below, the frequency as
well as the beam polarizations were chosen to null the AC Stark shift
from an individual beam and to produce a state-dependent force which
is equal in magnitude but opposite in sign for the $\left|\uparrow\right\rangle $
and $\left|\downarrow\right\rangle $ qubit states ($F_{\uparrow}=-F_{\downarrow}$
). The off-resonant laser beam frequency was detuned from the cycling
transition $\left(\left|\uparrow\right\rangle \rightarrow\left|P_{3/2},m_{J}=3/2\right\rangle \right)$
by $\Delta_{R}\simeq-63.8$ GHz. This gives detunings of $15.6$ GHz
and 26.1 GHz respectively from the $\left|\uparrow\right\rangle \rightarrow\left|P_{3/2},m_{J}=1/2\right\rangle $
and $\left|\downarrow\right\rangle \rightarrow\left|P_{3/2},m_{J}=-1/2\right\rangle $
transitions. Laser beam waists were $w_{z}\simeq100\:\mu$m in the
vertical (z-direction) and $w_{x}\simeq1$ mm in the horizontal direction.
Here we define the waist as the distance from the center of the beam
over which the electric field intensity decreases by $1/e^{2}$ (i.e.
$I(z)\sim e^{-(z/w_{z})^{2}}\:$). With the small $2.4^{o}$ incident
angle each beam makes with respect to the plane of the crystal, this
provided greater than 90\% uniform electric field intensity across
ion crystal arrays with $N<250$.

We used linearly polarized laser beams. Let
\[
\begin{array}{ccc}
\vec{E}_{U}\left(\vec{r},t\right) & = & \hat{\epsilon}_{U}E_{U}\cos\left(\vec{k}_{U}\cdot\vec{r}-\omega_{U}t\right)\\
\vec{E}_{L}\left(\vec{r},t\right) & = & \hat{\epsilon}_{L}E_{L}\cos\left(\vec{k}_{L}\cdot\vec{r}-\omega_{L}t\right)
\end{array}
\]
denote the electric fields of the upper and lower ODF beams. If $\phi_{p}$
is the angle of the laser beam electric-field polarization with respect
to vertical polarization $\left(\hat{\epsilon}\cdot\hat{x}=0\right)$,
then the AC Stark shift of the qubit states when illuminated by a
single beam can be written
\[
\begin{array}{c}
\Delta_{\uparrow,\, acss}=A_{\uparrow}\cos^{2}\left(\phi_{p}\right)+B_{\uparrow}\sin^{2}\left(\phi_{p}\right)\\
\Delta_{\downarrow,\, acss}=A_{\downarrow}\cos^{2}\left(\phi_{p}\right)+B_{\downarrow}\sin^{2}\left(\phi_{p}\right)
\end{array}
\]
where $A_{\uparrow}$($A_{\downarrow}$) is the Stark shift of the
$\left|\uparrow\right\rangle $($\left|\downarrow\right\rangle $)
state for a $\pi$-polarized beam ($\hat{\epsilon}$ parallel to the
$\hat{z}$-axis) and $B_{\uparrow}$($B_{\downarrow}$) is the Stark
shift of the $\left|\uparrow\right\rangle $($\left|\downarrow\right\rangle $)
state for a $\sigma$-polarized beam ($\hat{\epsilon}$ perpendicular
to the $\hat{z}$-axis). (Here we neglect the small $\sigma$ polarization
($\propto\sin(2.4^{o})$) that exists when $\phi_{p}=0$.) The Stark
shift of the qubit transition is
\begin{equation}
\Delta_{acss}=\left(A_{\uparrow}-A_{\downarrow}\right)\cos^{2}\left(\phi_{p}\right)+\left(B_{\uparrow}-B_{\downarrow}\right)\sin^{2}\left(\phi_{p}\right)\:.\label{eq:ACStark_shift}
\end{equation}
If $A_{\uparrow}-A_{\downarrow}$ and $B_{\uparrow}-B_{\downarrow}$
have opposite signs, there is an angle which makes $\Delta_{acss}=0$.
For a laser detuning of $\Delta_{R}=-63.8$ GHz, $\Delta_{acss}=0$
at $\phi_{p}\simeq\pm65{}^{o}$.\\

With $\Delta_{acss}=0$ for each ODF laser beam, we exploit the freedom
to choose their polarization in order to obtain a state-dependent
force. Specifically we choose $\vec{E}_{U}$ to have a polarization
given by $\phi_{p,u}=65{}^{o}$ and $\vec{E}_{L}$ to have a polarization
given by $\phi_{p,l}=-65{}^{o}$. In this case the interference term
in the expression for the electric field intensity $\left(\vec{E}_{U}+\vec{E}_{L}\right)^{2}$
produces a polarization gradient which results in spatially dependent
AC Stark shifts
\[
\begin{array}{c}
\left(A_{\uparrow}\cos^{2}\left(\phi_{p}\right)-B_{\uparrow}\sin^{2}\left(\phi_{p}\right)\right)2\sin\left(\delta k\cdot z+\mu_{R}t\right)\\
\left(A_{\downarrow}\cos^{2}\left(\phi_{p}\right)-B_{\downarrow}\sin^{2}\left(\phi_{p}\right)\right)2\sin\left(\delta k\cdot z+\mu_{R}t\right)
\end{array}
\]
for the qubit levels. Here $\delta k\equiv\left|\vec{k}_{U}-\vec{k}_{L}\right|=2k\sin\left(\frac{\theta_{R}}{2}\right)$
is the wave vector difference between the two ODF laser beams, $\mu_{R}=\omega_{U}-\omega_{L}$
is the ODF beat note, and $\phi_{p}=\left|\phi_{p,u}\right|=\left|\phi_{p,l}\right|$.
The spatially dependent AC Stark shift produces a state-dependent
force $F_{\uparrow,\downarrow}(z,t)=F_{\uparrow,\downarrow}\cos\left(\delta k\cdot z-\mu_{R}t\right)$
where
\[
\begin{array}{c}
F_{\uparrow}=2\,\delta k\left(A_{\uparrow}\cos^{2}\left(\phi_{p}\right)-B_{\uparrow}\sin^{2}\left(\phi_{p}\right)\right)\\
F_{\downarrow}=2\,\delta k\left(A_{\downarrow}\cos^{2}\left(\phi_{p}\right)-B_{\downarrow}\sin^{2}\left(\phi_{p}\right)\right)
\end{array}\:.
\]
In general $F_{\uparrow}\neq-F_{\downarrow}.$ We operate at $\Delta_{R}=-63.8$
GHz where for $\Delta_{acss}=0$ we also obtain $F_{\uparrow}=-F_{\downarrow}\equiv F$.

For a given $\phi_{p,u}\,$, $\phi_{p,l}\,$, and $\Delta_{R}$ we
use straight forward atomic physics along with well known values for
the energy levels and matrix elements of $^{9}$Be$^{+}$ to calculate
$F$ as a function of the electric field intensity $I_{R}=\frac{c\epsilon_{o}}{2}\left|E_{L}\right|^{2}=\frac{c\epsilon_{o}}{2}\left|E_{U}\right|^{2}$
at the center of the laser beams. For $\theta_{R}=4.8^{o}$ and $I_{R}=1\:$
W/cm$^{2}$ , $F=1.5\times10^{-23}$ N.

\section*{Wave Front Alignment}

\begin{figure}
\includegraphics[width=12cm]{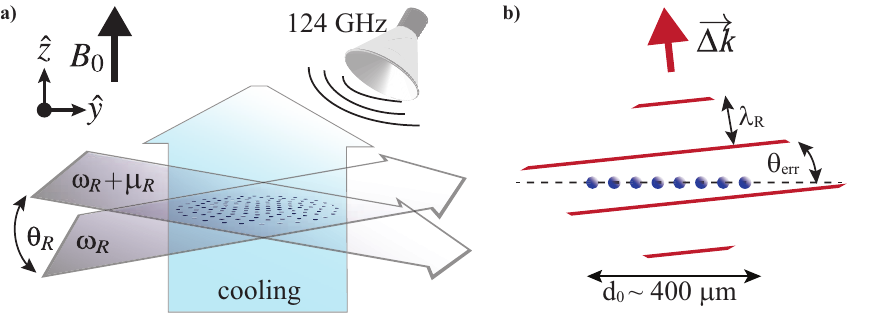}

\caption{\label{fig:tilted_wavefronts}Sketch of the 1D optical lattice wave
fronts (red lines) generated by the ODF laser beams. These wave fronts
need to be aligned with with the ion planar array (represented by
the blue dots). Here $\lambda_{R}=2\pi/|\protect\overrightarrow{\Delta k}|\approx3.7\,\mu\mbox{m}$
and $\theta_{err}$ denotes the angle of misalignment. $d_{0}\sim400\:\mu$m
is the typical array diameter for $N\sim200$ ions. With the wave
front alignment technique discussed in the text we obtain $\theta_{\mbox{err}}<0.05^{\circ}$.}
\end{figure}

\begin{figure}
\includegraphics[width=8cm]{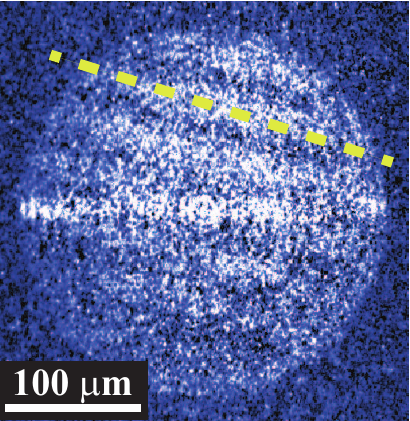}

\caption{\label{fig:wavefront_align}Top-view image of the spatially inhomogeneous
fluorescence from a single ion plane produced by the AC Stark from
a static ($\mu_{R}=0)$ optical dipole force lattice with misaligned
wave fronts. Dark bands are regions of high standing wave electric
field intensity (parallel to the dashed yellow line). The bright horizontal
feature bisecting the center of the image is fluorescence from the
weak Doppler laser cooling beam directed perpendicular to the magnetic
field. The image was obtained by subtracting a background image with
the ODF beams off.}
\end{figure}
The ODF laser beams produce a 1D optical lattice characterized by
the effective wave vector $\delta\vec{k}$ and beat note $\mu_{R}$.
In the previous section we assumed that $\delta\vec{k}\parallel\hat{z}$,
or equivalently that the wave fronts of the lattice were aligned perpendicular
to the $\hat{z}$-axis (magnetic field axis). If the wave fronts are
not normal to the $\hat{z}$-axis as sketched in Fig.~\ref{fig:tilted_wavefronts},
then the time dependence of the optical dipole force seen by an ion
in the rotating frame depends on the $(x,y)$ position of the ion.
This complicates the interaction generated by the optical dipole force
and is avoided by careful alignment.

We used top-view images (images of the ion resonance fluorescence
scattered along the magnetic field) from a single plane to measure
a misalignment of the ODF wave fronts. For this measurement we set
$\mu_{R}=0$ (stationary 1D lattice) and detune the frequency of the
ODF laser beams approximately 0.5 GHz below the $\left|\uparrow\right\rangle \rightarrow\left|^{2}P_{3/2}\: m_{J}=+3/2\right\rangle $
Doppler cooling transition. This small detuning generates sufficiently
large AC Stark shifts on the cooling transition to measurably change
the ion scatter rate from the Doppler cooling laser. With the Doppler
cooling laser on and the ODF beams turned off we observe a spatially
uniform, time-averaged image of a rotating planar crystal. With the
ODF beams on, ions located in regions of high electric field intensity
at the anti-nodes of the optical lattice are Stark shifted out of
resonance with the Doppler cooling laser. This is what produced the
dark bands in the top-view image shown in Fig.~\ref{fig:wavefront_align}.
From images like this we determine how to move the ODF beams to align
the wave fronts normal to $\hat{z}$. Improved alignment is indicated
by a longer wavelength fringe pattern. With this technique we have
aligned the ODF wave fronts with the planar array to better than $\theta_{err}\lesssim0.05^{o}$.

Images like that shown in Fig.~\ref{fig:wavefront_align} were typically
obtained with 1 s integration. This means the imprint of the 1D lattice
on the planar arrays was stable during the integration time and indicates
a phase stability of our 1D lattice of better than 1s. We note that
direct fluorescence imaging of the 1D lattice, for example by tuning
the ODF laser resonant with the Doppler cooling transition, is not
viable. Even at low powers, resonantly scattered photons across the
large horizontal waist of the ODF beams apply a large torque, causing
the rotation frequency and radius of the array to rapidly change,
typically driving the ions into very large radial orbits.

\section*{Spin-Motion Entanglement Produced by the Spin-Dependent Optical
Dipole Force}

With the wave vector $\delta\vec{k}$ of the 1D optical lattice aligned
parallel to $\hat{z}$, the optical dipole force generated by the
lattice is independent of the ion position and can be written
\begin{equation}
F_{\uparrow}(t)=-F_{\downarrow}(t)\equiv F\cos\left(\mu_{R}t\right)\label{eq:ODF force}
\end{equation}
where $\mu_{R}$ is the frequency difference between the ODF laser
beams. More generally we allow for the possibility that the ODF laser
intensity could be diferent for each ion, resulting in a different
spin-dependent force $F_{j}$ for each ion $j$,
\begin{equation}
F_{j\uparrow}(t)=-F_{j\downarrow}(t)\equiv F_{j}\cos\left(\mu_{R}t\right)\:.\label{eq:ODF force j}
\end{equation}
In the experimental set-up, the variation in $F_{j}$ is less than
20\%. The ODF interaction with the ion spins can be written as
\begin{equation}
H_{ODF}=-\sum_{j=1}^{N}F_{j}\cos\left(\mu_{R}t\right)\hat{z}_{j}\hat{\sigma}_{j}^{z}\:.\label{eq:ODF H}
\end{equation}
Here $\hat{z}_{j}$ is the axial position operator for the $j^{th}$
ion, which can be written in terms of the axial normal modes $\left(\vec{b}_{m},\omega_{m}\right)$
of the planar array,
\begin{equation}
\hat{z}_{j}=\sum_{m=1}^{N}b_{jm}\sqrt{\frac{\hslash}{2M\omega_{m}}}\left(\hat{a}_{m}e^{-i\omega_{m}t}+\hat{a}_{m}^{\dagger}e^{i\omega_{m}t}\right)\:.\label{eq:axial coord}
\end{equation}
The eigenvectors are normalized so that $\sum_{m}\left|b_{jm}\right|^{2}=\sum_{j}\left|b_{jm}\right|^{2}=1$.
Both the eigenvectors $\vec{b}_{m}$ and eigenfrequencies $\omega_{m}$
are calculated by solving for the ion equilibrium positions and diagonalizing
the stiffness matrix obtained by Taylor expansion of the potential
about the ion equilibrium positions~\cite{Britton11supp}.

The Hamiltonian $H_{ODF}$ of Eq.~(\ref{eq:ODF H}) is time dependent.
The evolution operator for $H_{ODF}$ is obtained from a second order
expansion of the Magnus formula~\cite{Zhu03,Kim09supp}
\begin{equation}
\hat{U}_{ODF}\left(t\right)=\exp\left[\frac{-i}{\hslash}\int_{0}^{t}H_{ODF}(t^{\prime})dt^{\prime}-\frac{1}{2\hslash^{2}}\int_{0}^{t}dt_{2}\int_{0}^{t_2}\left[H_{ODF}\left(t_{2}\right),H_{ODF}\left(t_{1}\right)\right]dt_{1}\right]\:.\label{eq:Magnus}
\end{equation}
Higher order terms do not contribute as the commutator $\left[H_{ODF}(t_{2}),H_{ODF}(t_{1})\right]$
commutes with $H_{ODF}(t^{\prime})$. Following the discussion of Ref.~\cite{Kim09supp}, $U_{ODF}(t)$ can be written
\begin{equation}
\begin{array}{ccc}
\hat{U}_{ODF}(t) & = & \exp\left[\sum_{j}\left(\sum_{m}\left(\alpha_{jm}(t)\hat{a}_{m}^{\dagger}-\alpha_{jm}^{\ast}(t)\hat{a}_{m}\right)\hat{\sigma}_{j}^{z}\right)+i\sum_{j,k}J_{j,k}(t)\hat{\sigma}_{j}^{z}\hat{\sigma}_{k}^{z}\right]\\
 & = & \exp\left[\sum_{j}\left(\sum_{m}\left(\alpha_{jm}(t)\hat{a}_{m}^{\dagger}-\alpha_{jm}^{\ast}(t)\hat{a}_{m}\right)\hat{\sigma}_{j}^{z}\right)\right]\cdot\exp\left[i\sum_{j,k}J_{j,k}(t)\hat{\sigma}_{j}^{z}\hat{\sigma}_{k}^{z}\right]\\
 & \equiv & \hat{U}_{SM}(t)\cdot\hat{U}_{SS}(t)
\end{array}\,.\label{eq:Uodf2}
\end{equation}
The first term $U_{SM}(t)$ describes spin-dependent displacements
$\alpha_{jm}(t)$ of the normal modes $m$ where, for the $\cos(\mu_{R}t)$
time dependence of the interaction in Eq.~(\ref{eq:ODF H}),
\begin{equation}
\alpha_{jm}(t)=\frac{F_{j}b_{jm}z_{0m}}{\hslash\left(\mu_{R}^{2}-\omega_{m}^{2}\right)}\left[\omega_{m}-e^{i\omega_{m}t}\left(\omega_{m}\cos(\mu_{R}t)-i\mu_{R}\sin(\mu_{R}t)\right)\right]\:.\label{eq:alpha_expr}
\end{equation}
Here $z_{0m}=\sqrt{\hslash/(2M\omega_{m})}$ . The second term $\hat{U}_{SS}(t)$
describes an effective spin-spin interaction where the pairwise coupling
$J_{j,k}(t)$ is given by
\begin{equation}
J_{j,k}(t)=\frac{F_{j}F_{k}}{2\hslash^{2}}\sum_{m}\frac{b_{jm}b_{km}z_{0m}^{2}}{\mu_{R}^{2}-\omega_{m}^{2}}\left\{ \frac{\omega_{m}\sin(\mu_{R}-\omega_{m})t}{\mu_{R}-\omega_{m}}+\frac{\omega_{m}\sin(\mu_{R}+\omega_{m})t}{\mu_{R}+\omega_{m}}-\frac{\omega_{m}\sin(2\mu_{R}t)}{2\mu_{R}}-\omega_{m}t\right\} \,.\label{eq:pairwise int coeff}
\end{equation}
For now we assume $\hat{U}_{SS}(t)$ can be neglected. We will discuss
the validity of this assumption at the end of this section.

The interaction $\hat{U}_{SM}(t)=\exp\left[\sum_{j}\left(\sum_{m}\left(\alpha_{jm}(t)\hat{a}_{m}^{\dagger}-\alpha_{jm}^{\ast}(t)\hat{a}_{m}\right)\hat{\sigma}_{j}^{z}\right)\right]$
generates spin-motion entanglement that is the subject of this study.
The commutator
\[
\begin{array}{ccc}
\left[\alpha_{jm}(t)\hat{a}_{m}^{\dagger}-\alpha_{jm}^{\ast}(t)\hat{a}_{m},\alpha_{km}(t)\hat{a}_{m}^{\dagger}-\alpha_{km}^{\ast}(t)\hat{a}_{m}\right] & = & \alpha_{jm}(t)\alpha_{km}^{\ast}(t)-\alpha_{jm}^{\ast}(t)\alpha_{km}(t)\\
 & = & 0
\end{array}
\]
because $\alpha_{jm}(t)\alpha_{km}^{\ast}(t)$ is real. Therefore
we can write $\hat{U}_{SM}(t)$ as a product of individual spin displacements
\begin{equation}
\hat{U}_{SM}(t)=\prod_{j,m}\exp\left(\left(\alpha_{jm}(t)\hat{a}_{m}^{\dagger}-\alpha_{jm}^{\ast}(t)\hat{a}_{m}\right)\hat{\sigma}_{j}^{z}\right)\:,\label{eq:USM_expression}
\end{equation}
which is Eq. (4) of the Letter. By neglecting the spin-spin
entanglement ($\hat{U}_{SS}(t)$) we can independently calculate the
evolution of each spin $j$.

\begin{figure}
\includegraphics[width=12cm]{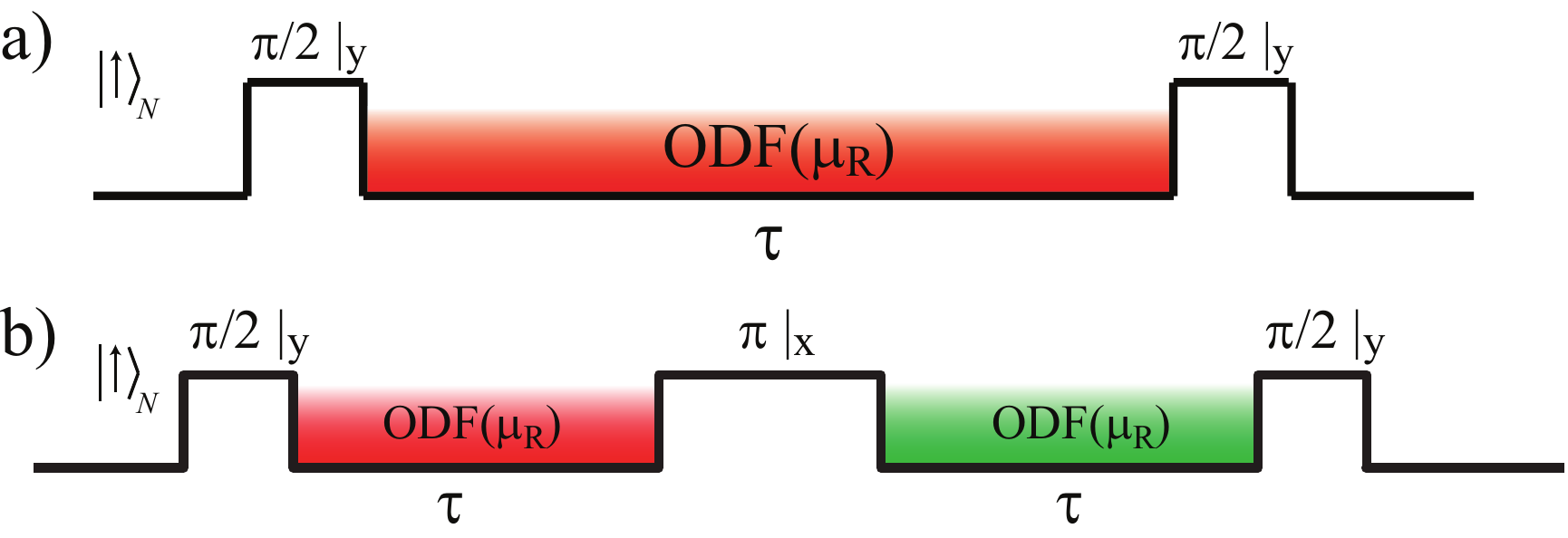}

\caption{Pulse sequences described in this supplemental material. a) Ramsey pulse sequence consisting of two $\pi/2$ rotations with an intermediate arm of duration $\tau$ during which the ODF is applied. b) The spin echo sequence repeated from Fig. 2(a) of the Letter which consists of two arms of duration $\tau$. \label{fig:Pulses}}
\end{figure}

We now calculate the spin motion entanglement generated by $\hat{U}_{SM}(t)$
during the free precession period of a Ramsey sequence shown in Fig.~\ref{fig:Pulses}(a). The calculation for the spin-echo sequence of Fig.~\ref{fig:Pulses}(b) used in the experiments
is identical except for a more complicated expression for the $\alpha_{jm}(t)^{\prime}$s
(see next section). Each spin $j$ is prepared in state $\left|\uparrow\right\rangle $
at the start of the sequence. If an ODF is not applied during the
free precession period, the spin is rotated to the dark $\left|\downarrow\right\rangle $
state by the final $\pi/2$ pulse of the sequence. With the application
of a spin-dependent ODF, in general the spin is entangled with the
motion at the end of the Ramsey sequence. We detect this spin-motion
entanglement by measuring the probability of finding spin $j$ in
the $\left|\uparrow\right\rangle $ state. Let
\[
\hat{U}_{SM}^{(j)}(t)=\exp\left(\sum_{m}\left(\alpha_{jm}(t)\hat{a}_{m}^{\dagger}-\alpha_{jm}^{\ast}(t)\hat{a}_{m}\right)\hat{\sigma}_{j}^{z}\right)
\]
denote the evolution of spin $j$ by the spin-dependent ODF. By re-writing
\[
\hat{U}_{SM}^{(j)}(t)=\cosh\left(\sum_{m}\left(\alpha_{jm}\hat{a}_{m}^{\dagger}-\alpha_{jm}^{\ast}\hat{a}_{m}\right)\right)+\sinh\left(\sum_{m}\left(\alpha_{jm}\hat{a}_{m}^{\dagger}-\alpha_{jm}^{\ast}\hat{a}_{m}\right)\right)\hat{\sigma}_{j}^{z}
\]
we calculate
\begin{equation}
P_{\uparrow,SM}^{(j)}=\left\langle \left(\sinh\left(\sum_{m}\left(\alpha_{jm}\hat{a}_{m}^{\dagger}-\alpha_{jm}^{\ast}\hat{a}_{m}\right)\right)\right)^{\dagger}\sinh\left(\sum_{m}\left(\alpha_{jm}\hat{a}_{m}^{\dagger}-\alpha_{jm}^{\ast}\hat{a}_{m}\right)\right)\right\rangle _{th}\label{eq:P_spin_up}
\end{equation}
where $P_{\uparrow,SM}^{(j)}$ denotes the probability of measuring
the $\left|\uparrow\right\rangle $ state for spin $j$ produced by
the $\hat{U}_{SM}(t)$ interaction, and $\left\langle \;\right\rangle _{th}$
denotes an expectation value averaged over a thermal (Maxwell-Boltzmann)
distribution of modes. We evaluate Eq.~(\ref{eq:P_spin_up}) by writing
the $\sinh$ functions in exponential form. It is then necessary to
evaluate expressions of the form $\left\langle e^{\hat{A}}e^{\hat{B}}\right\rangle _{th}$
where $\hat{A}$ and $\hat{B}$ are operators which are linear in
the raising and lowering operators $\hat{a}_{m}^{\dagger}$ and $\hat{a}_{m}$.
In this case we can make use of the result~\cite{AshcroftMermin}
\[
\left\langle e^{\hat{A}}e^{\hat{B}}\right\rangle _{th}=e^{(1/2)\left\langle \hat{A}^{2}+2\hat{A}\hat{B}+\hat{B}^{2}\right\rangle _{th}}
\]
to obtain
\begin{equation}
P_{\uparrow,SM}^{(j)}=\frac{1}{2}\left[1-\exp\left(-2\sum_{m}\left|\alpha_{jm}(t)\right|^{2}\left(2\bar{n}_{m}+1\right)\right)\right]\:.\label{eq:P_up_no_Gamma}
\end{equation}
Here $\bar{n}_{m}\simeq k_{B}T_{m}/(\hslash\omega_{m})$ is the mean
occupation number of a Maxwell-Boltzmann distribution characterized
by temperature $T_{m}$. We measure the probability of detecting $\left|\uparrow\right\rangle $
averaged over all the ions $\left(\sum_{j}P_{\uparrow}^{(j)}\right)/N$.\\

The simple result of Eq.~(\ref{eq:P_up_no_Gamma}) was obtained under
the assumption that we could neglect $\hat{U}_{SS}(t)$ in Eq.~(\ref{eq:Uodf2}).
In general $\hat{U}_{SS}(t)$ will contribute to the measured $P_{\uparrow}^{(j)}$.
This can be straight forwardly estimated when $\mu_{R}$ is tuned
close to the COM mode $\omega_{1}$. In this case the resulting pair-wise
interaction coefficients are identical for all ion pairs $J_{j,k}(t)\simeq J(t)$
with
\[
J(t)=\frac{F^{2}}{2\hslash^{2}}\cdot\frac{z_{01}^{2}}{N\left(\mu_{R}^{2}-\omega_{1}^{2}\right)}\left\{ \frac{\omega_{1}\sin(\mu_{R}-\omega_{1})t}{\mu_{R}-\omega_{1}}+\frac{\omega_{1}\sin(\mu_{R}+\omega_{1})t}{\mu_{R}+\omega_{1}}-\frac{\omega_{1}sin(2\mu_{R}t)}{2\mu_{R}}-\omega_{1}t\right\} \:.
\]
For small detunings $\left|\mu_{R}-\omega_{1}\right|\ll\omega_{1}$,
$J(t)$ is approximately bounded by $\left|J(t)\right|\lesssim J\cdot t$
where
\[
J=\frac{F^{2}}{2\hslash^{2}}\cdot\frac{z_{01}^{2}}{N\left(\mu_{R}^{2}-\omega_{1}^{2}\right)}\omega_{1}.
\]
The fully connected, uniform Ising interaction $\exp\left[iJ\left(\sum_{j,k}\hat{\sigma}_{j}^{z}\hat{\sigma}_{k}^{z}\right)t\right]$
obtained by coupling through the COM mode is identical to the single-axis
twisting interaction analyzed by Kitagawa and Ueda~\cite{Kitagawa93}. We use the expressions
given in Ref.~\cite{Kitagawa93}
to calculate $P_{\uparrow,SS}^{(j)}$, the probability of measuring
spin $j$ in the $\left|\uparrow\right\rangle $ state at the end
of the Ramsey sequence due to the $\hat{U}_{SS}(t)$ interaction,
\begin{equation}
P_{\uparrow,SS}^{(j)}\simeq\frac{1}{2}\left[N8(Jt)^{2}\right]\:.\label{eq:P_SS}
\end{equation}
This expression is valid for short times $t$ where $P_{\uparrow,SS}^{(j)}$
is small.

We obtain strong spin-motion entanglement for small detunings $\left|\mu_{R}-\omega_{1}\right|\ll\omega_{1}$.
The magnitude of the coherently driven amplitude $\alpha_{j,m=1}(t)$
in the expression for $\hat{U}_{SM}(t)$ (Eq.~(\ref{eq:USM_expression}))
and $P_{\uparrow,SM}^{(j)}$ (Eq.~(\ref{eq:P_up_no_Gamma})) is maximized
for a detuning $\left|\mu-\omega_{1}\right|\simeq\pi/t$ where
\[
\left|\alpha_{j,1}\right|_{max}=\left|\alpha_{j,1}\left(t\simeq\frac{\pi}{\left|\mu_{R}-\omega_{1}\right|}\right)\right|\simeq\frac{Fz_{01}}{\hslash\sqrt{N}\left|\mu_{R}^{2}-\omega_{1}^{2}\right|}2\omega_{1}\:.
\]
The above expression neglects terms of order $\left(\mu_{R}-\omega_{M}\right)/\omega_{M}$.
Inserting $\left|\alpha_{j,1}\right|_{max}$ into Eq.~(\ref{eq:P_up_no_Gamma})
and assuming the exponent is small gives
\begin{equation}
P_{\uparrow,SM}^{(j)}\simeq\frac{1}{2}\left[2\left|\alpha_{j,1}\right|_{max}^{2}\left(2\bar{n}_{1}+1)\right)\right]\:.\label{eq:P_SM}
\end{equation}
We compare $P_{\uparrow,SS}^{(j)}$ (Eq.~(\ref{eq:P_SS})) with $P_{\uparrow,SM}^{(j)}$
(Eq.~(\ref{eq:P_SM})),
\begin{equation}
\frac{P_{\uparrow,SS}^{(j)}}{P_{\uparrow,SM}^{(j)}}\simeq\frac{N\cdot8\left(Jt\right)^{2}}{2\left|\alpha_{j,1}\right|_{max}^{2}(2\bar{n}_{1}+1)}\simeq\frac{F^{2}}{4\hslash^{2}}\cdot\frac{z_{01}^{2}}{2\bar{n}_{1}+1}t^{2}\:.\label{eq:P_ratio}
\end{equation}
For the work reported here $F\sim10^{-23}$ N, $z_{01}=\sqrt{\hslash/\left(2M\omega_{1}\right)}\sim30$
nm, and $\bar{n}_{1}\sim10$ (Doppler cooling limit). For a typical
interaction time $t\lesssim10^{-3}$ s we calculate $P_{\uparrow,SS}^{(j)}/P_{\uparrow,SM}^{(j)}\lesssim0.1$.
Therefore for small detunings satisfying $\left|\mu_{R}-\omega_{1}\right|\lesssim(2\pi)/t\ll\omega_{1}$
we expect the spin-motion entanglement signature generated by $\hat{U}_{SM}(t)$
to dominate contributions due to $\hat{U}_{SS}(t)$. We note that
the spin-motion entanglement signature $\left(P_{\uparrow,SM}^{(j)}\right)$
decreases with temperature. For ground state cooling it may not be
possible to neglect $\hat{U}_{SS}(t)$.

We do not estimate $P_{\uparrow,SS}^{(j)}$ for $\mu_{R}$ tuned close
to modes $\omega_{M}$ other than the COM mode $\omega_{1}$. Therefore
we do not know if it is a good approximation to neglect $\hat{U}_{SS}(t)$
when resonantly coupling to non-COM modes. However, we experimentally
observe that neglecting $\hat{U}_{SS}(t)$ gives a good description
of our experimental measurements for $\mu_{R}$ tuned close to the
tilt ($\omega_{2}$ and $\omega_{3}$) and the next lower frequency
modes ($\omega_{4}$ and $\omega_{5}$).

\section*{Spin Echo Sequence with Decoherence}

To calculate $\alpha_{jm}(t)$ for the full spin echo sequence used in the experiment (see Fig.~\ref{fig:Pulses}(b)), we must account for the accumulated phase difference between the ODF drive and oscillating ion cloud over the first arm and intermediate microwave $\pi$-pulse of combined duration $(\tau+t_{\pi})$. This requires derivation of $\alpha_{jm}(t)$ for an ODF interaction with an arbitrary phase offset, $\phi$, given by the more general
\begin{equation}
H_{ODF}(\phi)=-\sum_{j=1}^{N}F_{j}\cos\left(\mu_{R}t+\phi\right)\hat{z}_{j}\hat{\sigma}_{j}^{z}, \label{eq:ODF Hphi}
\end{equation}
where $\phi = (\tau+t_{\pi})(\mu_R - \omega_m) = (\tau+t_{\pi})\delta_m$. Following the previous derivation of $\alpha_{jm}(t)$ for $\phi=0$ (Eq.~(\ref{eq:alpha_expr})), we obtain
\begin{equation}
\alpha_{jm}(t,\phi)=\frac{F_{j}b_{jm}z_{0m}}{\hslash\left(\mu_{R}^{2}-\omega_{m}^{2}\right)}\left[\omega_{m}\cos(\phi)-i\mu_R\sin(\phi)-e^{i\omega_{m}t}\left\{\omega_{m}\cos(\mu_{R}t+\phi)-i\mu_{R}\sin(\mu_{R}t+\phi)\right\}\right]\:.\label{eq:alpha_phi}
\end{equation}
We will now define a new $\alpha^{\text{SE}}_{jm}$ that may be substituted for $\alpha_{jm}$ in Eq.~(\ref{eq:USM_expression}) to calculate $P_{\uparrow}$ for the full spin echo sequence exhibiting arm durations of $\tau$:
\begin{equation}
\alpha^{\text{SE}}_{jm} = \alpha_{jm}(\tau,\phi=0) - \alpha_{jm}(\tau,\phi), \label{eq:alpha_se}
\end{equation}
where the above expression is given explicitly in Eq. (6) of the Letter.

To justify implementation of Eq.~(\ref{eq:alpha_se}), it is useful to calculate $P_{\uparrow}$ for a single spin undergoing both the Ramsey and spin echo sequences. To simplify notation, we define the displacement operator $\hat{D}(\alpha_{jm})=\exp(\alpha_{jm} \hat{a}^{\dag}_m - \alpha^{\ast}_{jm}\hat{a}_m)$ which is applied separately to $|\!\!\uparrow_j\rangle\otimes|\psi_m\rangle$ and $|\!\!\downarrow_j\rangle\otimes|\psi_m\rangle$, where $|\psi_m\rangle$ is an arbitrary motional state of mode $m$. Assuming the state is initialized to $|\!\!\uparrow_j\rangle\otimes|\psi_m\rangle$, we calculate the result of the Ramsey sequence, $P_{\uparrow}^{(j)\text{Ramsey}}$, to be
\begin{equation}
P_{\uparrow}^{(j)\text{Ramsey}} = \frac{1}{4} \langle\psi_m|\left\{\hat{D}^{\dag}\left(\alpha_{jm}(\tau,\phi)\right)-\hat{D}^{\dag}\left(-\alpha_{jm}(\tau,\phi)\right)\right\}\left\{h.c.\right\}|\psi_m\rangle \label{eq:Pup_Ram},
\end{equation}
where $\left\{h.c.\right\}$ denotes the Hermitian conjugate of the first bracketed expression. Here the arbitrary phase $\phi$ has no physical significance since its value is common to all displacements, and we have once again made the assumption that $F_{j\uparrow} = -F_{j\downarrow}$. However, the spin echo result given by $P^{(j)\text{SE}}_{\uparrow}$ is
\begin{eqnarray}
P^{(j)\text{SE}}_{\uparrow} &=& \frac{1}{4} \langle\psi_m|\left\{\hat{D}^{\dag}\left(-\alpha_{jm}(\tau,\phi)\right)\hat{D}^{\dag}\left(\alpha_{jm}(\tau,0)\right)-\hat{D}^{\dag}\left(\alpha_{jm}(\tau,\phi)\right)\hat{D}^{\dag}\left(-\alpha_{jm}(\tau,0)\right)\right\}\left\{h.c.\right\}|\psi_m\rangle \label{eq:Pse1}\\
& = & \frac{1}{4} \langle\psi_m|\left\{\hat{D}^{\dag}\left(\alpha^{\text{SE}}_{jm}\right)-\hat{D}^{\dag}\left(-\alpha^{\text{SE}}_{jm}\right)\right\}\left\{h.c.\right\}|\psi_m\rangle. \label{eq:Pse2}
\end{eqnarray}
We obtain Eq.~(26) from Eq.~(25) using the multiplicative properties of $\hat{D}$ and neglecting overall phase factors that leave $P^{(j)\text{SE}}_{\uparrow}$ unchanged. Note that Eq.~(26) is identical to Eq.~(24) after an appropriate redefinition of $\alpha_{jm}$.

Finally, the derivation of Eq.~(\ref{eq:P_up_no_Gamma}) neglected
the effects of spontaneous emission from the ODF laser beams. Decoherence
of the Bloch vector due to spontaneous emission from off-resonant
light is well studied in our system~\cite{Uys10supp}.
The qubit levels are closed under spontaneous light scattering; that
is, spontaneous light scattering does not optically pump an ion to
a different ground state level outside of the two qubit levels. In
the presence of off-resonant laser light, the decrease in the Bloch
vector due to spontaneous scattering during the arms of a spin-echo
sequence is
\[
P_{\uparrow,spon}^{(j)}=\frac{1}{2}\left[1-\exp\left(-\Gamma\cdot2\tau\right)\right]\:.
\]
Here $\Gamma\equiv\left(\Gamma_{Ram}+\Gamma_{el}\right)/2$ has contributions
from both Raman scattering and elastic Rayleigh scattering that can
be calculated from the laser beam parameters. With the spin echo sequence, we account for spontaneous
emission by modifying Eq.~(\ref{eq:P_up_no_Gamma}) as follows
\[
P_{\uparrow,SM}^{(j)\text{SE}}=\frac{1}{2}\left[1-e^{-\Gamma 2\tau}\exp\left(-2\sum_{m}\left|\alpha_{jm}^{\text{SE}}\right|^{2}\left(2\bar{n}_{m}+1\right)\right)\right],
\]
where $\tau$ is the length of time of a single arm of the spin-echo sequence.




\begin{thebibliography}{39}
\expandafter\ifx\csname natexlab\endcsname\relax\def\natexlab#1{#1}\fi
\expandafter\ifx\csname bibnamefont\endcsname\relax
  \def\bibnamefont#1{#1}\fi
\expandafter\ifx\csname bibfnamefont\endcsname\relax
  \def\bibfnamefont#1{#1}\fi
\expandafter\ifx\csname citenamefont\endcsname\relax
  \def\citenamefont#1{#1}\fi
\expandafter\ifx\csname url\endcsname\relax
  \def\url#1{\texttt{#1}}\fi
\expandafter\ifx\csname urlprefix\endcsname\relax\def\urlprefix{URL }\fi
\providecommand{\bibinfo}[2]{#2}
\providecommand{\eprint}[2][]{\url{#2}}

\bibitem[{\citenamefont{O'Connell et~al.}(2010)}]{OConnell10}
\bibinfo{author}{\bibfnamefont{A.~D.} \bibnamefont{O'Connell}}
  \bibnamefont{et~al.}, \bibinfo{journal}{Nature}
  \textbf{\bibinfo{volume}{464}}, \bibinfo{pages}{697} (\bibinfo{year}{2010}).

\bibitem[{\citenamefont{Kippenberg and Vahala}(2008)}]{Kippenberg08}
\bibinfo{author}{\bibfnamefont{T.~J.} \bibnamefont{Kippenberg}}
  \bibnamefont{and} \bibinfo{author}{\bibfnamefont{K.~J.}
  \bibnamefont{Vahala}}, \bibinfo{journal}{Science}
  \textbf{\bibinfo{volume}{321}}, \bibinfo{pages}{1172} (\bibinfo{year}{2008}).

\bibitem[{\citenamefont{Teufel et~al.}(2009)\citenamefont{Teufel, Donner,
  Castellanos-Beltran, Harlow, and Lehnert}}]{Teufel09}
\bibinfo{author}{\bibfnamefont{J.~D.} \bibnamefont{Teufel}},
  \bibinfo{author}{\bibfnamefont{T.}~\bibnamefont{Donner}},
  \bibinfo{author}{\bibfnamefont{M.~A.} \bibnamefont{Castellanos-Beltran}},
  \bibinfo{author}{\bibfnamefont{J.~W.} \bibnamefont{Harlow}},
  \bibnamefont{and} \bibinfo{author}{\bibfnamefont{K.~W.}
  \bibnamefont{Lehnert}}, \bibinfo{journal}{Nat. Nanotechnol.}
  \textbf{\bibinfo{volume}{4}}, \bibinfo{pages}{820} (\bibinfo{year}{2009}).

\bibitem[{\citenamefont{Biercuk et~al.}(2010)\citenamefont{Biercuk, Britton,
  Uys, VanDevender, and Bollinger}}]{Biercuk10}
\bibinfo{author}{\bibfnamefont{M.~J.} \bibnamefont{Biercuk}},
  \bibinfo{author}{\bibfnamefont{J.~W.} \bibnamefont{Britton}},
  \bibinfo{author}{\bibfnamefont{H.}~\bibnamefont{Uys}},
  \bibinfo{author}{\bibfnamefont{A.}~\bibnamefont{VanDevender}},
  \bibnamefont{and} \bibinfo{author}{\bibfnamefont{J.~J.}
  \bibnamefont{Bollinger}}, \bibinfo{journal}{Nat. Nanotechnol.}
  \textbf{\bibinfo{volume}{5}}, \bibinfo{pages}{646} (\bibinfo{year}{2010}).

\bibitem[{\citenamefont{Jost et~al.}(2009)\citenamefont{Jost, Home, Amini,
  Hanneke, Ozeri, Langer, Bollinger, Leibfried, and Wineland}}]{Jost09}
\bibinfo{author}{\bibfnamefont{J.~D.} \bibnamefont{Jost}},
  \bibinfo{author}{\bibfnamefont{J.~P.} \bibnamefont{Home}},
  \bibinfo{author}{\bibfnamefont{J.~M.} \bibnamefont{Amini}},
  \bibinfo{author}{\bibfnamefont{D.}~\bibnamefont{Hanneke}},
  \bibinfo{author}{\bibfnamefont{R.}~\bibnamefont{Ozeri}},
  \bibinfo{author}{\bibfnamefont{C.}~\bibnamefont{Langer}},
  \bibinfo{author}{\bibfnamefont{J.~J.} \bibnamefont{Bollinger}},
  \bibinfo{author}{\bibfnamefont{D.}~\bibnamefont{Leibfried}},
  \bibnamefont{and} \bibinfo{author}{\bibfnamefont{D.~J.}
  \bibnamefont{Wineland}}, \bibinfo{journal}{Nature}
  \textbf{\bibinfo{volume}{459}}, \bibinfo{pages}{683} (\bibinfo{year}{2009}).

\bibitem[{\citenamefont{Brown et~al.}(2011)\citenamefont{Brown, Ospelkaus,
  Colombe, Wilson, and Wineland}}]{Brown11}
\bibinfo{author}{\bibfnamefont{K.~R.} \bibnamefont{Brown}},
  \bibinfo{author}{\bibfnamefont{C.}~\bibnamefont{Ospelkaus}},
  \bibinfo{author}{\bibfnamefont{Y.}~\bibnamefont{Colombe}},
  \bibinfo{author}{\bibfnamefont{A.~C.} \bibnamefont{Wilson}},
  \bibnamefont{and} \bibinfo{author}{\bibfnamefont{D.~J.}
  \bibnamefont{Wineland}}, \bibinfo{journal}{Nature}
  \textbf{\bibinfo{volume}{471}}, \bibinfo{pages}{196} (\bibinfo{year}{2011}).

\bibitem[{\citenamefont{Ichimaru}(1982)}]{Ichimaru82}
\bibinfo{author}{\bibfnamefont{S.}~\bibnamefont{Ichimaru}},
  \bibinfo{journal}{Rev. Mod. Phys.} \textbf{\bibinfo{volume}{54}},
  \bibinfo{pages}{1017} (\bibinfo{year}{1982}).

\bibitem[{\citenamefont{Dubin and O'Neil}(1999)}]{Dubin99}
\bibinfo{author}{\bibfnamefont{D.~H.~E.} \bibnamefont{Dubin}} \bibnamefont{and}
  \bibinfo{author}{\bibfnamefont{T.~M.} \bibnamefont{O'Neil}},
  \bibinfo{journal}{Rev. Mod. Phys.} \textbf{\bibinfo{volume}{71}},
  \bibinfo{pages}{87} (\bibinfo{year}{1999}).

\bibitem[{\citenamefont{Hanneke et~al.}(2009)\citenamefont{Hanneke, Home,
  Amini, Leibfried, and Wineland}}]{Hanneke09}
\bibinfo{author}{\bibfnamefont{D.}~\bibnamefont{Hanneke}},
  \bibinfo{author}{\bibfnamefont{J.~P.} \bibnamefont{Home}},
  \bibinfo{author}{\bibfnamefont{J.~M.} \bibnamefont{Amini}},
  \bibinfo{author}{\bibfnamefont{D.}~\bibnamefont{Leibfried}},
  \bibnamefont{and} \bibinfo{author}{\bibfnamefont{D.~J.}
  \bibnamefont{Wineland}}, \bibinfo{journal}{Nature Phys.}
  \textbf{\bibinfo{volume}{6}}, \bibinfo{pages}{13} (\bibinfo{year}{2009}).

\bibitem[{\citenamefont{Monz et~al.}(2009)}]{Monz09}
\bibinfo{author}{\bibfnamefont{T.}~\bibnamefont{Monz}} \bibnamefont{et~al.},
  \bibinfo{journal}{Phys. Rev. Lett.} \textbf{\bibinfo{volume}{103}},
  \bibinfo{pages}{200503} (\bibinfo{year}{2009}).

\bibitem[{\citenamefont{Friedenauer et~al.}(2008)\citenamefont{Friedenauer,
  Schmitz, Glueckert, Porras, and Schaetz}}]{Friedenauer08}
\bibinfo{author}{\bibfnamefont{A.}~\bibnamefont{Friedenauer}},
  \bibinfo{author}{\bibfnamefont{H.}~\bibnamefont{Schmitz}},
  \bibinfo{author}{\bibfnamefont{J.~T.} \bibnamefont{Glueckert}},
  \bibinfo{author}{\bibfnamefont{D.}~\bibnamefont{Porras}}, \bibnamefont{and}
  \bibinfo{author}{\bibfnamefont{T.}~\bibnamefont{Schaetz}},
  \bibinfo{journal}{Nature Phys.} \textbf{\bibinfo{volume}{4}},
  \bibinfo{pages}{757} (\bibinfo{year}{2008}).

\bibitem[{\citenamefont{Kim et~al.}(2010)\citenamefont{Kim, Chang, Korenblit,
  Islam, Edwards, Freericks, Lin, Duan, and Monroe}}]{Kim10}
\bibinfo{author}{\bibfnamefont{K.}~\bibnamefont{Kim}},
  \bibinfo{author}{\bibfnamefont{M.-S.} \bibnamefont{Chang}},
  \bibinfo{author}{\bibfnamefont{S.}~\bibnamefont{Korenblit}},
  \bibinfo{author}{\bibfnamefont{R.}~\bibnamefont{Islam}},
  \bibinfo{author}{\bibfnamefont{E.~E.} \bibnamefont{Edwards}},
  \bibinfo{author}{\bibfnamefont{J.~K.} \bibnamefont{Freericks}},
  \bibinfo{author}{\bibfnamefont{G.-D.} \bibnamefont{Lin}},
  \bibinfo{author}{\bibfnamefont{L.-M.} \bibnamefont{Duan}}, \bibnamefont{and}
  \bibinfo{author}{\bibfnamefont{C.}~\bibnamefont{Monroe}},
  \bibinfo{journal}{Nature} \textbf{\bibinfo{volume}{465}},
  \bibinfo{pages}{590} (\bibinfo{year}{2010}).

\bibitem[{\citenamefont{Islam et~al.}(2011)}]{Islam11}
\bibinfo{author}{\bibfnamefont{R.}~\bibnamefont{Islam}} \bibnamefont{et~al.},
  \bibinfo{journal}{Nat. Commun.} \textbf{\bibinfo{volume}{2}},
  \bibinfo{pages}{377} (\bibinfo{year}{2011}).

\bibitem[{\citenamefont{Britton et~al.}(2012)}]{Britton11}
\bibinfo{author}{\bibfnamefont{J.~W.} \bibnamefont{Britton}}
  \bibnamefont{et~al.}, \bibinfo{journal}{Nature} \textbf{\bibinfo{volume}{484}},
  \bibinfo{pages}{489} (\bibinfo{year}{2012}).

\bibitem[{\citenamefont{Lanyon et~al.}(2011)}]{Lanyon11}
\bibinfo{author}{\bibfnamefont{B.~P.} \bibnamefont{Lanyon}}
  \bibnamefont{et~al.}, \bibinfo{journal}{Science}
  \textbf{\bibinfo{volume}{334}}, \bibinfo{pages}{57} (\bibinfo{year}{2011}).

\bibitem[{\citenamefont{Biercuk
  et~al.}(2009{\natexlab{a}})\citenamefont{Biercuk, Uys, VanDevender, Shiga,
  Itano, and Bollinger}}]{Biercuk09nature}
\bibinfo{author}{\bibfnamefont{M.~J.} \bibnamefont{Biercuk}},
  \bibinfo{author}{\bibfnamefont{H.}~\bibnamefont{Uys}},
  \bibinfo{author}{\bibfnamefont{A.~P.} \bibnamefont{VanDevender}},
  \bibinfo{author}{\bibfnamefont{N.}~\bibnamefont{Shiga}},
  \bibinfo{author}{\bibfnamefont{W.~M.} \bibnamefont{Itano}}, \bibnamefont{and}
  \bibinfo{author}{\bibfnamefont{J.~J.} \bibnamefont{Bollinger}},
  \bibinfo{journal}{Nature} \textbf{\bibinfo{volume}{458}},
  \bibinfo{pages}{996} (\bibinfo{year}{2009}{\natexlab{a}}).

\bibitem[{\citenamefont{Rosenband et~al.}(2008)}]{Rosenband08}
\bibinfo{author}{\bibfnamefont{T.}~\bibnamefont{Rosenband}}
  \bibnamefont{et~al.}, \bibinfo{journal}{Science}
  \textbf{\bibinfo{volume}{319}}, \bibinfo{pages}{1808} (\bibinfo{year}{2008}).

\bibitem[{\citenamefont{Heinzen et~al.}(1991)\citenamefont{Heinzen, Bollinger,
  Moore, Itano, and Wineland}}]{Heinzen91}
\bibinfo{author}{\bibfnamefont{D.~J.} \bibnamefont{Heinzen}},
  \bibinfo{author}{\bibfnamefont{J.~J.} \bibnamefont{Bollinger}},
  \bibinfo{author}{\bibfnamefont{F.~L.} \bibnamefont{Moore}},
  \bibinfo{author}{\bibfnamefont{W.~M.} \bibnamefont{Itano}}, \bibnamefont{and}
  \bibinfo{author}{\bibfnamefont{D.~J.} \bibnamefont{Wineland}},
  \bibinfo{journal}{Phys. Rev. Lett.} \textbf{\bibinfo{volume}{66}},
  \bibinfo{pages}{2080} (\bibinfo{year}{1991}).

\bibitem[{\citenamefont{Bollinger et~al.}(1993)\citenamefont{Bollinger,
  Heinzen, Moore, Itano, Wineland, and Dubin}}]{Bollinger93}
\bibinfo{author}{\bibfnamefont{J.~J.} \bibnamefont{Bollinger}},
  \bibinfo{author}{\bibfnamefont{D.~J.} \bibnamefont{Heinzen}},
  \bibinfo{author}{\bibfnamefont{F.~L.} \bibnamefont{Moore}},
  \bibinfo{author}{\bibfnamefont{W.~M.} \bibnamefont{Itano}},
  \bibinfo{author}{\bibfnamefont{D.~J.} \bibnamefont{Wineland}},
  \bibnamefont{and} \bibinfo{author}{\bibfnamefont{D.~H.~E.}
  \bibnamefont{Dubin}}, \bibinfo{journal}{Phys. Rev. A}
  \textbf{\bibinfo{volume}{48}}, \bibinfo{pages}{525} (\bibinfo{year}{1993}).

\bibitem[{\citenamefont{Weimer et~al.}(1994)\citenamefont{Weimer, Bollinger,
  Moore, and Wineland}}]{Weimer94}
\bibinfo{author}{\bibfnamefont{C.~S.} \bibnamefont{Weimer}},
  \bibinfo{author}{\bibfnamefont{J.~J.} \bibnamefont{Bollinger}},
  \bibinfo{author}{\bibfnamefont{F.~L.} \bibnamefont{Moore}}, \bibnamefont{and}
  \bibinfo{author}{\bibfnamefont{D.~J.} \bibnamefont{Wineland}},
  \bibinfo{journal}{Phys. Rev. A} \textbf{\bibinfo{volume}{49}},
  \bibinfo{pages}{3842} (\bibinfo{year}{1994}).

\bibitem[{\citenamefont{Tinkle et~al.}(1994)\citenamefont{Tinkle, Greaves,
  Surko, Spencer, and Mason}}]{Tinkle94}
\bibinfo{author}{\bibfnamefont{M.~D.} \bibnamefont{Tinkle}},
  \bibinfo{author}{\bibfnamefont{R.~G.} \bibnamefont{Greaves}},
  \bibinfo{author}{\bibfnamefont{C.~M.} \bibnamefont{Surko}},
  \bibinfo{author}{\bibfnamefont{R.~L.} \bibnamefont{Spencer}},
  \bibnamefont{and} \bibinfo{author}{\bibfnamefont{G.~W.} \bibnamefont{Mason}},
  \bibinfo{journal}{Phys. Rev. Lett.} \textbf{\bibinfo{volume}{72}},
  \bibinfo{pages}{352} (\bibinfo{year}{1994}).

\bibitem[{\citenamefont{Dantan et~al.}(2010)\citenamefont{Dantan, Marler,
  Albert, Gu\'{e}not, and Drewsen}}]{Dantan10}
\bibinfo{author}{\bibfnamefont{A.}~\bibnamefont{Dantan}},
  \bibinfo{author}{\bibfnamefont{J.~P.} \bibnamefont{Marler}},
  \bibinfo{author}{\bibfnamefont{M.}~\bibnamefont{Albert}},
  \bibinfo{author}{\bibfnamefont{D.}~\bibnamefont{Gu\'{e}not}},
  \bibnamefont{and} \bibinfo{author}{\bibfnamefont{M.}~\bibnamefont{Drewsen}},
  \bibinfo{journal}{Phys. Rev. Lett.} \textbf{\bibinfo{volume}{105}},
  \bibinfo{pages}{103001} (\bibinfo{year}{2010}).

\bibitem[{\citenamefont{Kriesel et~al.}(2002)\citenamefont{Kriesel, Bollinger,
  Mitchell, King, and Dubin}}]{Kriesel02}
\bibinfo{author}{\bibfnamefont{J.~M.} \bibnamefont{Kriesel}},
  \bibinfo{author}{\bibfnamefont{J.~J.} \bibnamefont{Bollinger}},
  \bibinfo{author}{\bibfnamefont{T.~B.} \bibnamefont{Mitchell}},
  \bibinfo{author}{\bibfnamefont{L.~B.} \bibnamefont{King}}, \bibnamefont{and}
  \bibinfo{author}{\bibfnamefont{D.~H.~E.} \bibnamefont{Dubin}},
  \bibinfo{journal}{Phys. Rev. Lett.} \textbf{\bibinfo{volume}{88}},
  \bibinfo{pages}{125003} (\bibinfo{year}{2002}).

\bibitem[{\citenamefont{Castro et~al.}(2010)\citenamefont{Castro, McQuillen,
  and Killian}}]{Castro10}
\bibinfo{author}{\bibfnamefont{J.}~\bibnamefont{Castro}},
  \bibinfo{author}{\bibfnamefont{P.}~\bibnamefont{McQuillen}},
  \bibnamefont{and} \bibinfo{author}{\bibfnamefont{T.~C.}
  \bibnamefont{Killian}}, \bibinfo{journal}{Phys. Rev. Lett.}
  \textbf{\bibinfo{volume}{105}}, \bibinfo{pages}{065004}
  (\bibinfo{year}{2010}).

\bibitem[{\citenamefont{Jensen et~al.}(2004)\citenamefont{Jensen, Hasegawa, and
  Bollinger}}]{Jensen04}
\bibinfo{author}{\bibfnamefont{M.~J.} \bibnamefont{Jensen}},
  \bibinfo{author}{\bibfnamefont{T.}~\bibnamefont{Hasegawa}}, \bibnamefont{and}
  \bibinfo{author}{\bibfnamefont{J.~J.} \bibnamefont{Bollinger}},
  \bibinfo{journal}{Phys. Rev. A} \textbf{\bibinfo{volume}{70}},
  \bibinfo{pages}{033401} (\bibinfo{year}{2004}).

\bibitem[{\citenamefont{Monroe et~al.}(1995)}]{Monroe95}
\bibinfo{author}{\bibfnamefont{C.}~\bibnamefont{Monroe}} \bibnamefont{et~al.},
  \bibinfo{journal}{Phys. Rev. Lett.} \textbf{\bibinfo{volume}{75}},
  \bibinfo{pages}{4011} (\bibinfo{year}{1995}).

\bibitem[{\citenamefont{Biercuk
  et~al.}(2009{\natexlab{b}})\citenamefont{Biercuk, Uys, VanDevender, Shiga,
  Itano, and Bollinger}}]{Biercuk09}
\bibinfo{author}{\bibfnamefont{M.~J.} \bibnamefont{Biercuk}},
  \bibinfo{author}{\bibfnamefont{H.}~\bibnamefont{Uys}},
  \bibinfo{author}{\bibfnamefont{A.~P.} \bibnamefont{VanDevender}},
  \bibinfo{author}{\bibfnamefont{N.}~\bibnamefont{Shiga}},
  \bibinfo{author}{\bibfnamefont{W.~M.} \bibnamefont{Itano}}, \bibnamefont{and}
  \bibinfo{author}{\bibfnamefont{J.~J.} \bibnamefont{Bollinger}},
  \bibinfo{journal}{Quantum Info. and Comp.} \textbf{\bibinfo{volume}{9}},
  \bibinfo{pages}{920} (\bibinfo{year}{2009}{\natexlab{b}}).

\bibitem[{\citenamefont{Hasegawa et~al.}(2005)\citenamefont{Hasegawa, Jensen,
  and Bollinger}}]{Hasegawa05}
\bibinfo{author}{\bibfnamefont{T.}~\bibnamefont{Hasegawa}},
  \bibinfo{author}{\bibfnamefont{M.~J.} \bibnamefont{Jensen}},
  \bibnamefont{and} \bibinfo{author}{\bibfnamefont{J.~J.}
  \bibnamefont{Bollinger}}, \bibinfo{journal}{Phys. Rev. A}
  \textbf{\bibinfo{volume}{71}}, \bibinfo{pages}{023406}
  (\bibinfo{year}{2005}).

\bibitem[{\citenamefont{Huang et~al.}(1998)\citenamefont{Huang, Bollinger,
  Mitchell, Itano, and Dubin}}]{Huang98}
\bibinfo{author}{\bibfnamefont{X.-P.} \bibnamefont{Huang}},
  \bibinfo{author}{\bibfnamefont{J.~J.} \bibnamefont{Bollinger}},
  \bibinfo{author}{\bibfnamefont{T.~B.} \bibnamefont{Mitchell}},
  \bibinfo{author}{\bibfnamefont{W.~M.} \bibnamefont{Itano}}, \bibnamefont{and}
  \bibinfo{author}{\bibfnamefont{D.~H.~E.} \bibnamefont{Dubin}},
  \bibinfo{journal}{Phys. Plasmas} \textbf{\bibinfo{volume}{5}},
  \bibinfo{pages}{1656} (\bibinfo{year}{1998}).

\bibitem[{\citenamefont{Mitchell et~al.}(1998)\citenamefont{Mitchell,
  Bollinger, Dubin, Huang, Itano, and Baughman}}]{Mitchell98}
\bibinfo{author}{\bibfnamefont{T.~B.} \bibnamefont{Mitchell}},
  \bibinfo{author}{\bibfnamefont{J.~J.} \bibnamefont{Bollinger}},
  \bibinfo{author}{\bibfnamefont{D.~H.~E.} \bibnamefont{Dubin}},
  \bibinfo{author}{\bibfnamefont{X.-P.} \bibnamefont{Huang}},
  \bibinfo{author}{\bibfnamefont{W.~M.} \bibnamefont{Itano}}, \bibnamefont{and}
  \bibinfo{author}{\bibfnamefont{R.~H.} \bibnamefont{Baughman}},
  \bibinfo{journal}{Science} \textbf{\bibinfo{volume}{282}},
  \bibinfo{pages}{1290} (\bibinfo{year}{1998}).

\bibitem[{\citenamefont{Zhu et~al.}(2006)\citenamefont{Zhu, Monroe, and
  Duan}}]{Zhu06}
\bibinfo{author}{\bibfnamefont{S.-L.} \bibnamefont{Zhu}},
  \bibinfo{author}{\bibfnamefont{C.}~\bibnamefont{Monroe}}, \bibnamefont{and}
  \bibinfo{author}{\bibfnamefont{L.-M.} \bibnamefont{Duan}},
  \bibinfo{journal}{Phys. Rev. Lett.} \textbf{\bibinfo{volume}{97}},
  \bibinfo{pages}{050505} (\bibinfo{year}{2006}).

\bibitem[{\citenamefont{Kim et~al.}(2009)\citenamefont{Kim, Chang, Islam,
  Korenblit, Duan, and Monroe}}]{Kim09}
\bibinfo{author}{\bibfnamefont{K.}~\bibnamefont{Kim}},
  \bibinfo{author}{\bibfnamefont{M.-S.} \bibnamefont{Chang}},
  \bibinfo{author}{\bibfnamefont{R.}~\bibnamefont{Islam}},
  \bibinfo{author}{\bibfnamefont{S.}~\bibnamefont{Korenblit}},
  \bibinfo{author}{\bibfnamefont{L.-M.} \bibnamefont{Duan}}, \bibnamefont{and}
  \bibinfo{author}{\bibfnamefont{C.}~\bibnamefont{Monroe}},
  \bibinfo{journal}{Phys. Rev. Lett.} \textbf{\bibinfo{volume}{103}},
  \bibinfo{pages}{120502} (\bibinfo{year}{2009}).

\bibitem[{EPA()}]{EPAPS}
\bibinfo{note}{See Supplemental Material for technical details and
  derivations}.

\bibitem[{\citenamefont{Hahn}(1950)}]{Hahn50}
\bibinfo{author}{\bibfnamefont{E.~L.} \bibnamefont{Hahn}},
  \bibinfo{journal}{Phys. Rev.} \textbf{\bibinfo{volume}{80}},
  \bibinfo{pages}{580} (\bibinfo{year}{1950}).

\bibitem[{\citenamefont{Uys et~al.}(2009)\citenamefont{Uys, Biercuk, and
  Bollinger}}]{Uys09}
\bibinfo{author}{\bibfnamefont{H.}~\bibnamefont{Uys}},
  \bibinfo{author}{\bibfnamefont{M.~J.} \bibnamefont{Biercuk}},
  \bibnamefont{and} \bibinfo{author}{\bibfnamefont{J.~J.}
  \bibnamefont{Bollinger}}, \bibinfo{journal}{Phys. Rev. Lett.}
  \textbf{\bibinfo{volume}{103}}, \bibinfo{pages}{040501}
  (\bibinfo{year}{2009}).

\bibitem[{\citenamefont{Monroe et~al.}(1996)\citenamefont{Monroe, Meekhoff,
  King, and Wineland}}]{Monroe96}
\bibinfo{author}{\bibfnamefont{C.}~\bibnamefont{Monroe}},
  \bibinfo{author}{\bibfnamefont{D.~M.} \bibnamefont{Meekhoff}},
  \bibinfo{author}{\bibfnamefont{B.~E.} \bibnamefont{King}}, \bibnamefont{and}
  \bibinfo{author}{\bibfnamefont{D.~J.} \bibnamefont{Wineland}},
  \bibinfo{journal}{Science} \textbf{\bibinfo{volume}{272}},
  \bibinfo{pages}{1131} (\bibinfo{year}{1996}).

\bibitem[{\citenamefont{Uys et~al.}(2010)\citenamefont{Uys, Biercuk,
  VanDevender, Ospelkaus, Meiser, Ozeri, and Bollinger}}]{Uys10}
\bibinfo{author}{\bibfnamefont{H.}~\bibnamefont{Uys}},
  \bibinfo{author}{\bibfnamefont{M.~J.} \bibnamefont{Biercuk}},
  \bibinfo{author}{\bibfnamefont{A.}~\bibnamefont{VanDevender}},
  \bibinfo{author}{\bibfnamefont{C.}~\bibnamefont{Ospelkaus}},
  \bibinfo{author}{\bibfnamefont{D.}~\bibnamefont{Meiser}},
  \bibinfo{author}{\bibfnamefont{R.}~\bibnamefont{Ozeri}}, \bibnamefont{and}
  \bibinfo{author}{\bibfnamefont{J.~J.} \bibnamefont{Bollinger}},
  \bibinfo{journal}{Phys. Rev. Lett.} \textbf{\bibinfo{volume}{105}},
  \bibinfo{pages}{200401} (\bibinfo{year}{2010}).

\bibitem[{\citenamefont{Porras and Cirac}(2004)}]{Porras04}
\bibinfo{author}{\bibfnamefont{D.}~\bibnamefont{Porras}} \bibnamefont{and}
  \bibinfo{author}{\bibfnamefont{J.~I.} \bibnamefont{Cirac}},
  \bibinfo{journal}{Phys. Rev. Lett.} \textbf{\bibinfo{volume}{92}},
  \bibinfo{pages}{207901} (\bibinfo{year}{2004}).

\bibitem[{\citenamefont{Porras and Cirac}(2006)}]{Porras06}
\bibinfo{author}{\bibfnamefont{D.}~\bibnamefont{Porras}} \bibnamefont{and}
  \bibinfo{author}{\bibfnamefont{J.~I.} \bibnamefont{Cirac}},
  \bibinfo{journal}{Phys. Rev. Lett.} \textbf{\bibinfo{volume}{96}},
  \bibinfo{pages}{250501} (\bibinfo{year}{2006}).

\end{thebibliography}

\begin{thebibliography}{6}
\expandafter\ifx\csname natexlab\endcsname\relax\def\natexlab#1{#1}\fi
\expandafter\ifx\csname bibnamefont\endcsname\relax
  \def\bibnamefont#1{#1}\fi
\expandafter\ifx\csname bibfnamefont\endcsname\relax
  \def\bibfnamefont#1{#1}\fi
\expandafter\ifx\csname citenamefont\endcsname\relax
  \def\citenamefont#1{#1}\fi
\expandafter\ifx\csname url\endcsname\relax
  \def\url#1{\texttt{#1}}\fi
\expandafter\ifx\csname urlprefix\endcsname\relax\def\urlprefix{URL }\fi
\providecommand{\bibinfo}[2]{#2}
\providecommand{\eprint}[2][]{\url{#2}}

\bibitem[{\citenamefont{Britton et~al.}(2011)}]{Britton11supp}
\bibinfo{author}{\bibfnamefont{J.~W.} \bibnamefont{Britton}}
  \bibnamefont{et~al.}, \bibinfo{journal}{Nature} \textbf{\bibinfo{volume}{484}},
  \bibinfo{pages}{489} (\bibinfo{year}{2012}).

\bibitem[{\citenamefont{Zhu and Wang}(2003)}]{Zhu03}
\bibinfo{author}{\bibfnamefont{S.-L.} \bibnamefont{Zhu}} \bibnamefont{and}
  \bibinfo{author}{\bibfnamefont{Z.~D.} \bibnamefont{Wang}},
  \bibinfo{journal}{Phys. Rev. Lett.} \textbf{\bibinfo{volume}{91}},
  \bibinfo{pages}{187902} (\bibinfo{year}{2003}).

\bibitem[{\citenamefont{Kim et~al.}(2009)\citenamefont{Kim, Chang, Islam,
  Korenblit, Duan, and Monroe}}]{Kim09supp}
\bibinfo{author}{\bibfnamefont{K.}~\bibnamefont{Kim}},
  \bibinfo{author}{\bibfnamefont{M.-S.} \bibnamefont{Chang}},
  \bibinfo{author}{\bibfnamefont{R.}~\bibnamefont{Islam}},
  \bibinfo{author}{\bibfnamefont{S.}~\bibnamefont{Korenblit}},
  \bibinfo{author}{\bibfnamefont{L.-M.} \bibnamefont{Duan}}, \bibnamefont{and}
  \bibinfo{author}{\bibfnamefont{C.}~\bibnamefont{Monroe}},
  \bibinfo{journal}{Phys. Rev. Lett.} \textbf{\bibinfo{volume}{103}},
  \bibinfo{pages}{120502} (\bibinfo{year}{2009}).

\bibitem[{\citenamefont{Ashcroft and Mermin}(1976)}]{AshcroftMermin}
\bibinfo{author}{\bibfnamefont{N.~W.} \bibnamefont{Ashcroft}} \bibnamefont{and}
  \bibinfo{author}{\bibfnamefont{N.~D.} \bibnamefont{Mermin}},
  \emph{\bibinfo{title}{Solid State Physics}} (\bibinfo{publisher}{Saunders
  College, Philadelphia}, \bibinfo{year}{1976}).

\bibitem[{\citenamefont{Kitagawa and Ueda}(1993)}]{Kitagawa93}
\bibinfo{author}{\bibfnamefont{M.}~\bibnamefont{Kitagawa}} \bibnamefont{and}
  \bibinfo{author}{\bibfnamefont{M.}~\bibnamefont{Ueda}},
  \bibinfo{journal}{Phys. Rev. A} \textbf{\bibinfo{volume}{47}},
  \bibinfo{pages}{5138} (\bibinfo{year}{1993}).

\bibitem[{\citenamefont{Uys et~al.}(2010)\citenamefont{Uys, Biercuk,
  VanDevender, Ospelkaus, Meiser, Ozeri, and Bollinger}}]{Uys10supp}
\bibinfo{author}{\bibfnamefont{H.}~\bibnamefont{Uys}},
  \bibinfo{author}{\bibfnamefont{M.~J.} \bibnamefont{Biercuk}},
  \bibinfo{author}{\bibfnamefont{A.}~\bibnamefont{VanDevender}},
  \bibinfo{author}{\bibfnamefont{C.}~\bibnamefont{Ospelkaus}},
  \bibinfo{author}{\bibfnamefont{D.}~\bibnamefont{Meiser}},
  \bibinfo{author}{\bibfnamefont{R.}~\bibnamefont{Ozeri}}, \bibnamefont{and}
  \bibinfo{author}{\bibfnamefont{J.~J.} \bibnamefont{Bollinger}},
  \bibinfo{journal}{Phys. Rev. Lett.} \textbf{\bibinfo{volume}{105}},
  \bibinfo{pages}{200401} (\bibinfo{year}{2010}).

\end{thebibliography}

\end{document}